\documentclass[12pt,preprint]{aastex3}
\usepackage{epsfig}
\newcommand{\vvec}{{\mathbf v}}

\newcommand{\uvec}{{\mathbf u}}

\newcommand{\bvec}{{\mathbf B}}

\newcommand{\svec}{{\mathbf R}}
\newcommand{\hatn}{\,\hat{\mathbf n}}

\newcommand{\be}{\begin{equation}}
\newcommand{\ee}{\end{equation}}
\newcommand{\bea}{\begin{eqnarray}}
\newcommand{\eea}{\end{eqnarray}}
\newcommand{\beax}{\begin{eqnarray*}}
\newcommand{\eeax}{\end{eqnarray*}}
\newcommand{\ba}{\begin{array}}
\newcommand{\ea}{\end{array}}
\newcommand{\bed}{\begin{description}}
\newcommand{\ed}{\end{description}}
\newcommand{\blc}{\begin{list}{$\circ$}{}}
\newcommand{\blb}{\begin{list}{$\bullet$}{}}
\newcommand{\el}{\end{list}}
\newcommand{\ben}{\begin{enumerate}}
\newcommand{\een}{\end{enumerate}}

\def\lapprox{\mathrel{\hbox{\rlap{\hbox{\lower4pt\hbox{$\sim$}}}\hbox{$<$}}}}
\def\gapprox{\mathrel{\hbox{\rlap{\hbox{\lower4pt\hbox{$\sim$}}}\hbox{$>$}}}}

\begin{document}

\title{Decorrelation Times of Photospheric Fields and Flows}

\author{B.~T. Welsch}
\affil{Space Sciences Laboratory, University of California, 
Berkeley, CA 94720-7450}

\author{K. Kusano\altaffilmark{1}}
\affil{Solar Terrestrial Environment Laboratory (STEL), Nagoya University}
\altaffiltext{1}{Japan Agency for Marine-Earth Science and Technology
(JAMSTEC),Yokohama, Kanagawa 236-0001, Japan}

\author{T. T. Yamamoto}
\affil{STEL, Nagoya University}

\author{K. Muglach} 
\affil{Code 674, NASA Goddard Space Flight Center,
  Greenbelt, MD 20771; Also at ARTEP, Inc., Ellicott City, MD 21042}

\begin{abstract}
We use autocorrelation to investigate evolution in flow fields inferred by applying Fourier Local Correlation Tracking (FLCT) to a sequence of high-resolution (0.3 \arcsec), high-cadence ($\simeq 2$ min) line-of-sight magnetograms of NOAA active region (AR) 10930 recorded by the Narrowband Filter Imager (NFI) of the Solar Optical Telescope (SOT) aboard the {\em Hinode} satellite over 12--13 December 2006. To baseline the timescales of flow evolution, we also autocorrelated the magnetograms, at several spatial binnings, to characterize the lifetimes of active region magnetic structures versus spatial scale. Autocorrelation of flow maps can be used to optimize tracking parameters, to understand tracking algorithms' susceptibility to noise, and to estimate flow lifetimes. Tracking parameters varied include: time interval $\Delta t$ between magnetogram pairs tracked, spatial binning applied to the magnetograms, and windowing parameter $\sigma$ used in FLCT. Flow structures vary over a range of spatial and temporal scales (including unresolved scales), so tracked flows represent a local average of the flow over a particular range of space and time. We define flow lifetime to be the flow decorrelation time, $\tau$. For $\Delta t > \tau$, tracking results represent the average velocity over one or more flow lifetimes. We analyze lifetimes of flow components, divergences, and curls as functions of magnetic field strength and spatial scale. We find a significant trend of increasing lifetimes of flow components, divergences, and curls with field strength, consistent with Lorentz forces partially governing flows in the active photosphere, as well as strong trends of increasing flow lifetime and decreasing magnitudes with increases in both spatial scale and $\Delta t$.
\end{abstract}

\section{Introduction}
\label{sec:intro}

Estimates of photospheric velocities can be combined with the ideal
Ohm's law to determine the fluxes of magnetic energy and helicity
across the photosphere (e.g., \citealt{Chae2001, Demoulin2003,
  Schuck2006}).  These fluxes probably play important roles in driving
the corona to flare \citep{Welsch2009} and to launch coronal mass
ejections (CMEs), as well as coronal heating \citep{Tan2007}.  Several
techniques can be used to estimate photospheric flows, including
tracking methods (e.g., \citealt{Fisher2008, Schuck2006}),
balltracking \citep{Potts2004} and feature tracking
\citep{DeForest2007}.  Techniques for estimating photospheric
velocities, however, are imperfect \citep{Rieutord2001, Welsch2007,
  Schuck2008}, so efforts to improve estimation techniques are
ongoing.

Autocorrelation of image sequences has been used to quantify
evolutionary timescales of structures in the solar atmosphere.  In
this approach, a measure of correlation (e.g., linear or rank-order
correlation) is computed between pixel values at some initial time
$t_i$ and the corresponding pixel values at a later time $t_i + \Delta
T$, where we refer to $\Delta T$ as the lag time.  As image features
evolve with increasing $\Delta T$, measures of correlation computed
this way typically drop.  For instance, \citet{Tritschler2007}
investigated internetwork and network fine structure of the quiet Sun
at disk center in the Ca II K line.  \citet{Welsch2009} autocorrelated
96-minute-cadence, line-of-sight (LOS) magnetograms of a few dozen
active regions imaged by the MDI instrument \citep{Scherrer1995} in
full-disk mode (with $\simeq 2 \arcsec$ pixels) over lag times ranging
from the data's nominal 96-minute cadence to several days.  They found
decorrelation times for the magnetic field, defined as a drop in
the autocorrelation to $1/e \sim 0.37$, on the order of $72$ hours,
with considerable variation between active regions. \citet{Welsch2009}
also estimated horizontal flows in their active region sample by
applying tracking methods to pairs of magnetograms separated in time
by a tracking interval, $\Delta t$, equal to the data's 96-minute
cadence.  The flows were independently estimated by using two distinct
methods, Fourier local correlation tracking (FLCT;
\citealt{Fisher2008}) and the differential affine velocity estimator
(DAVE; \citealt{Schuck2006}).  Apodization windows of 8 and 9 pixels,
respectively, were used, so motions on smaller scales were averaged
away over the apodization window.  By autocorrelating the $x$- and
$y$-components of the estimated flows, they found flow lifetimes of
two to three 96-minute steps on this $\sim 12$ Mm spatial scale.

The autocorrelation approach employed by \citet{Welsch2009} was useful
for estimating the lifetime of magnetic field structures and flows on
the particular spatial scales they studied.  But what are the
lifetimes of magnetic and flow structures on different spatial scales?
How do flow lifetimes vary as a function of length scale?  And what
are the lifetimes of flow properties, such as vortical or converging
motions?  From the general properties of turbulent flows like those
operating at the solar photosphere, we expect structure in flow fields
to be present over a range of spatial scales, with a range of
lifetimes.  

Further, little effort has been undertaken to understand how
variations in tracking cadence $\Delta t$, noise, and choice
apodization window size affect flow inference. How do variations in
these quantities influence estimated flows?  \citet{Abramenko2011}
used feature tracking to follow bright points over a range of $\Delta
t$'s, and characterized their motion as super-diffusive.  The average
speeds implied by their diffusive model decreased with increasing
$\Delta t$.  \citet{Chae2000} applied an LCT algorithm to
chromospheric H-$\alpha$ images, and noted that $\Delta t$ must be
long enough for displacements to be large enough to be detected by
LCT, but not so long that displacements are larger than the
apodization window.  They also found that choosing window sizes that
were too small increased random velocities due to noise.
\citet{Chae2004b} found slower average speeds for longer $\Delta t$'s.
More recently, \citet{Verma2011} conducted an investigation of the
effects of parameter selection on flows inferred by applying local
correlation tracking (LCT) to high-cadence, high-resolution (0.11
\arcsec pixel scale) G-band images of a field of view containing
sunspot umbrae, penumbrae, and quiet-Sun regions, observed by the
broadband filter imager (BFI) of the SOT \citep{Tsuneta2008} aboard
the {\em Hinode} satellite \citep{Kosugi2007}.  The evolutionary time
scales and spatial scales of the underlying image structures motivated
their selection of tracking intervals between 15 and 200 s, and
apodization windows of 600 -- 2400 km.  They found decreases in flow
speeds both as the apodization window size increased, and as flow maps
were averaged over longer time periods.  Because photospheric magnetic
flux is long-lived, tracking methods can be applied to magnetograms
separated in time by much larger tracking intervals --- e.g., longer
than an hour, as done by \citet{Welsch2009}.

Here, we address these and other topics by using FLCT to track a
nearly 13-hour sequence of high-resolution ($0.16 \arcsec$ pixels, with a
$\sim 0.32 \arcsec$ diffraction limit), rapid cadence ($\simeq 2$ min) LOS
magnetograms of NOAA AR 10930, recorded by the SOT/NFI
\citep{Tsuneta2008, Suematsu2008, Ichimoto2008, Shimizu2008}
instrument aboard {\em Hinode} over 2006 December 12 -- 13.  This active
region was the source of an X-class flare and CME during our tracking
interval: the flare began around 13-Dec-2006 02:15UT and peaked around
02:40 UT in GOES 1--8 \AA \, 
X-rays.
\citet{Tan2009} also investigated flows in this active region,
focusing on the relationship of penumbral flows and evolution to the
flare, and \citet{Schrijver2008} conducted a comparative study of
coronal magnetic field extrapolations derived from vector magnetograms
of this active region.

This paper's primary goals are: first, to investigate the effects of
choices of tracking parameters on inferred flows; and second, to
characterize the lifetimes and other properties of flows as functions
of spatial scale and magnetic field.  We begin by describing the
magnetogram dataset in more detail (\S \ref{sec:data}), then describe
the methods we used to estimate flows (\S \ref{sec:methods}), present
our analysis of the lifetimes of fields (\S \ref{sec:b_lifetimes}) and
flows (\S \ref{sec:flow_lifetimes}), and conclude with a brief summary
of our results and their implications (\S \ref{sec:discussion}).


\section{Hinode SOT/NFI Magnetograms}
\label{sec:data}

\begin{figure}[ht]
  \centerline{%
\psfig{figure=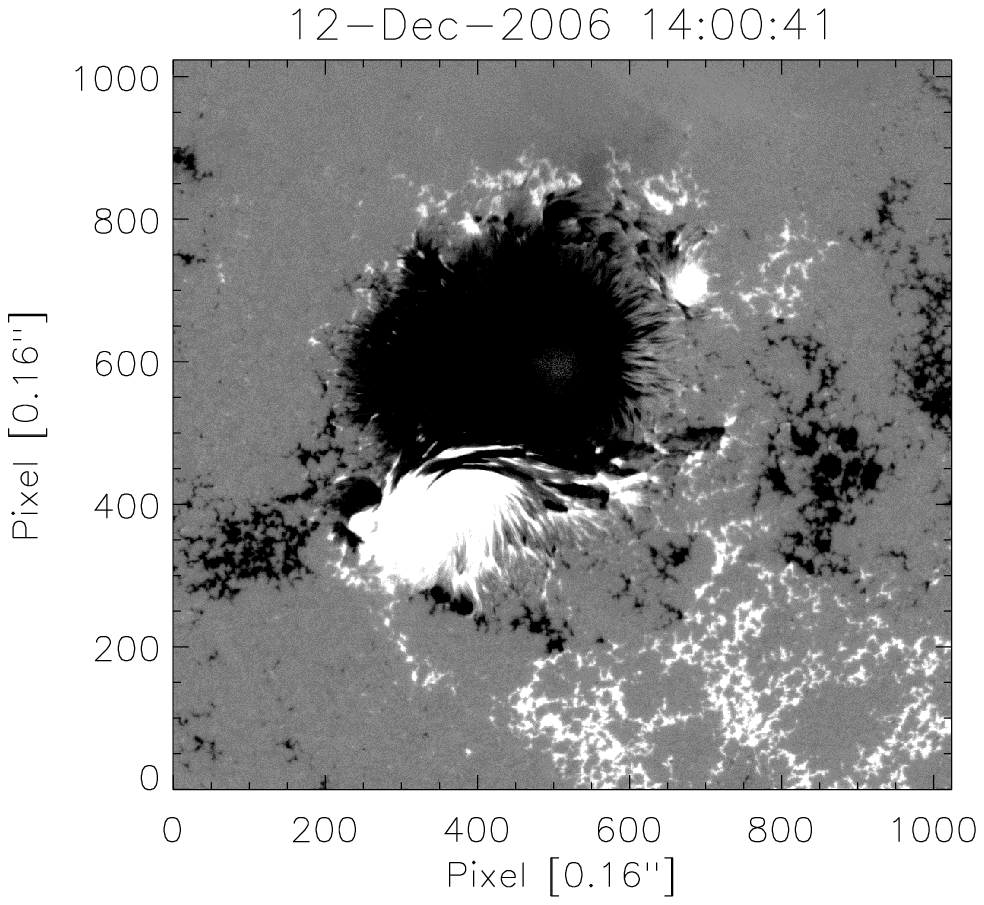,width=3.25in} %
\psfig{figure=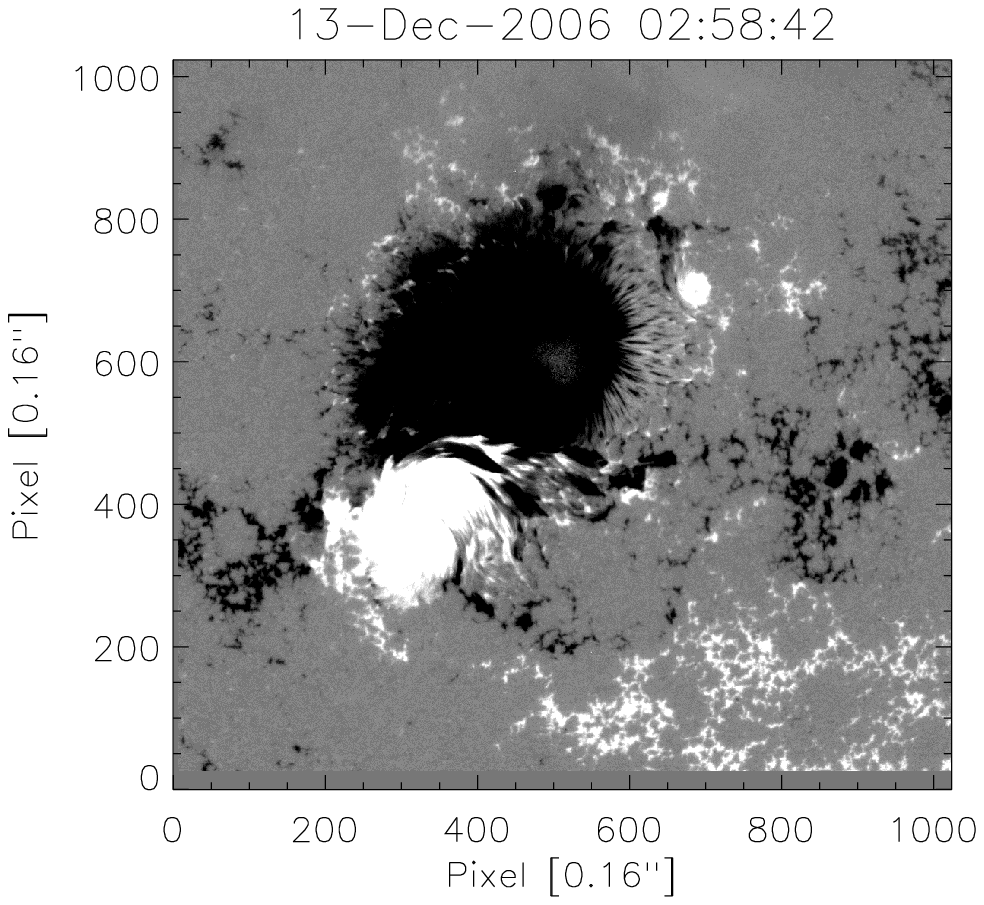,width=3.25in}}   
  \caption[]{\footnotesize \textsl{Left: Initial Hinode/NFI
      LOS magnetogram (white is positive flux, black is negative)
      from the 13-hour data sequence we investigated, at full
      resolution (0.16 \arcsec pixels) with the saturation level set at $\pm
      500$ Mx cm$^{-2}$.  Right: Final magnetogram in the sequence.}}
        \label{fig:nfi_i_f}
\end{figure}

We studied a sequence of {\em Hinode} SOT/NFI Fe I 6302 \AA shuttered
magnetograms of AR 10930 with $0.16 \arcsec$ pixels, created from the
Stokes $V/I$ ratio in Level 0 data, recorded between 12-Dec-2006 at
14:00 and 13-Dec-2006 at 02:58.  The USAF/NOAA Solar Region Summary
issued at 24:00 UT on 12-Dec-2006 listed AR 10930 at S06W21, meaning
it was near disk center during this interval.  At this wavelength, the
diffraction limit of the Solar Optical Telescope's 50 cm aperture is
$1.22 \lambda / a \simeq 0.32 \arcsec$.  We show initial and final
magnetograms from the sequence in Figure \ref{fig:nfi_i_f}.  At that
stage of the {\em Hinode} mission, a bubble present within the NFI
instrument degraded image quality in part of the field of view. In
these frames, the affected pixels are in an area between from about
300 -- 900 in $x$ and above about $900$ in $y$.  To account for the
bubble, we exclude pixels above $y = 848$ in all our analyses of flow
properties and lifetimes (although this region is retained in some
images shown in this paper).

To convert the measured Stokes $I$ and $V$ signals into pixel-averaged
flux densities, we applied the approximate calibration used by
\citet{Isobe2007},
\be B_{\rm LOS} = - \frac{C_V}{0.798 C_I} \times 10000 ~~ ({\rm Mx \, cm}^{-2}) 
~, \ee \label{eqn:isobe}
\noindent
where $B_{\rm LOS}$ is the estimated LOS flux density, $C_I$ and $C_V$
are counts in the $I$ and $V$ images, respectively, and $C_V/0.798$
gives the circular polarization.  As noted by \citet{Isobe2007}, this
linear scaling is only approximate.  In particular, in the dark cores
of umbrae, the decrease in $I$ introduces nonlinearity into the $V/I$
ratio. This effect can produce spurious weak field regions in umbrae.
A more complex, nonlinear calibration would be required to overcome
this artifact.  Since our primary goal is to investigate the time
evolution of structure in this magnetogram sequence, we have not
pursued efforts to correct such artifacts in the $V/I$ magnetograms.

A detailed spectroscopic analysis of fields in this active region
using {\em Hinode} SpectroPolarimeter (SP; \citealt{Tsuneta2008}) data was
undertaken by \cite{Schrijver2008}, who created one vector magnetogram
with approximately 0.63 \arcsec pixels near the middle of our tracking
interval, around 21:00 UT.%
%
%
Reprojection employed in their reduction procedure limits the ability
to directly apply those results with our magnetogram data set, but, as
described in Appendix \ref{app:sp}, we manually co-registered and
resampled this SP vector magnetogram to map vertical magnetic field
$B_z$ and field strength $|\bvec|$ from this magnetogram onto the $(4
\times 4)$-binned NFI magnetogram closest to the midpoint of the
45-minute SP scan (at 21:52 UT).  

The AR was at about S06W21 toward the end of our observations, and
NFI's field of view (FOV) subtends about 10 heliocentric degrees at
disk center.  These facts imply cosines between the line of sight and
the local vertical direction within the AR ranging from 0.88 to 0.96,
corresponding to variations in pixel scales from foreshortening of a
few percent across the FOV.  Over the earlier part of magnetogram
sequence, upon which much of our analysis is focused, these variations
are slightly smaller.  Consequently, we take these modest distortions
as acceptable, and opt not to reproject the magnetograms prior to
tracking.  Reprojection is, however, certainly appropriate for
tracking studies over much larger fields of view, e.g., that of
\citet{Welsch2009}, who estimated velocities simultaneously across
fields of view out to 45$^\circ$ from disk center.

Apart from three abnormally large time steps of approximately 600 s
each, and two relatively small time steps of 26 s each, the 380 frames
we analyze have a cadence of 121.4 $\pm$ 1.2 s.

%
%

%

\begin{figure}[ht]
  \centerline{\psfig{figure=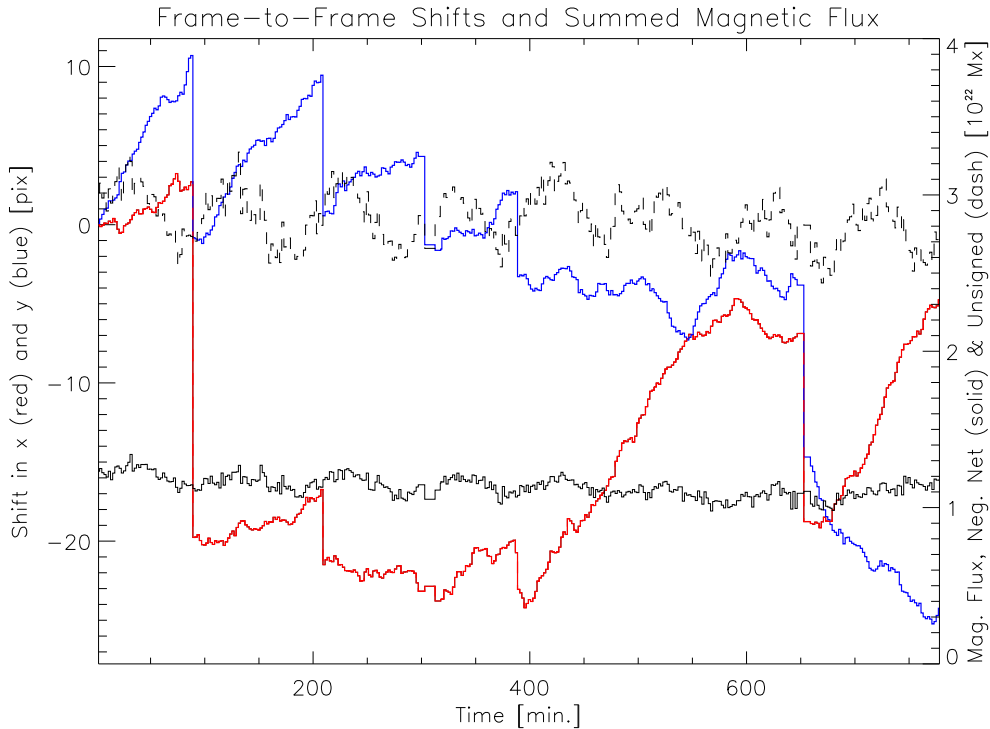,width=5.5in}}   
  \caption[]{\footnotesize \textsl{ Left axis: Cumulative
      frame-to-frame shifts (in pixels) in the $x$ (red) and $y$
      (blue) directions between pairs of magnetograms in the NFI
      sequence that we analyzed, with respect to the initial
      magnetogram. Right axis: total unsigned flux (dashed) and
      the negative of net flux (solid).  Variations in flux
      probably arise from crosstalk due to Doppler shifts induced by
      spacecraft orbital motion into the Stokes $V$ signal.}}
        \label{fig:shifts}
\end{figure}

To study frame-to-frame changes in the magnetic field, we must first
remove artificial evolution due to whole-image shifts (rigid motion of
each image) between frames, due primarily to variations in spacecraft
pointing.  To accomplish this, we first determined the frame-to-frame
shift between the central regions of each pair of images (cropping the
outer 10\% of pixels along each edge), using FFTs to compute the
frame-to-frame cross correlation function and a second-order Taylor
series approximation to estimate the displacement vector of the
function's peak, to sub-pixel accuracy \citep{Fisher2008}.  We then
used Fourier interpolation to estimate pixel values implied by a shift
of the negative of the cumulative displacement from the initial frame
in the sequence.  Any wrapped pixels arising from this shift, due to
the assumed periodicity, were then zeroed (as may be seen in a thin
strip along the bottom of the right panel of Figure
\ref{fig:nfi_i_f}).  In Figure \ref{fig:shifts}, we show the
cumulative shifts in $x$ and $y$ over the duration of the magnetogram
sequence.  Large shifts in image alignment occur with an approximate
periodicity similar to that of Hinode's 98-minute orbital period. %
\citep{Katsukawa2010}.

In Figure \ref{fig:shifts}, we also show the total unsigned flux and
negative of the net flux (there is more negative LOS flux than
positive; but the right plot axis is positive, so we changed the sign
of the net flux), which both exhibit periodicities on a timescale
similar to that of the whole-frame shifts.  Analysis of Fourier
spectra of the time series in pixels (not shown) shows a significant
excess power near the orbital frequency, but little evidence of
helioseismic P-mode leakage into the magnetogram signal.  Given the
large Doppler shift from orbital motion, observed periodicities could
arise from leakage of Doppler shifts induced by spacecraft orbital
motion into the Stokes $V$ measurement.  Thermal variations in the
instrument might also be responsible.

As discussed in more detail below, local fluctuations in field
strength from noise can mimic changes from true magnetic evolution,
implying estimated flows will be inaccurate in regions where the
measured fields are near the magnetograph noise level.  We estimated
this noise level using the procedure of \citet{Hagenaar1999}.  First,
we spatially binned the data $2 \times 2$ to $0.32 \arcsec$-wide
pixels, to approximate the SOT diffraction limit of $0.3 \arcsec$.
Next, we created histograms of pixel values in each frame in the
sequence, in bins of 3 Mx cm$^{-2}$.  Then, assuming the core of the
distribution in signals arises purely from noise, we fit a Gaussian,
$\exp[-(B-B_0)^2/(2 a^2)$, to the central region of the histogram,
  $\pm 10$ Mx cm$^{-2}$.  Over the course of the tracking interval,
  fitted values of the width parameter $a$ for single magnetograms
  ranged from a high near 17 Mx cm$^{-2}$ at the start of the run to
  lows near 7 Mx cm$^{-2}$ toward the end of the run.  For simplicity,
  we therefore take 15 Mx cm$^{-2}$ as a uniform estimate of the noise
  level.

\section{Estimating Flows}
\label{sec:methods}

Assuming magnetic diffusivity is negligibly small over the time
interval $\Delta t$ between a pair of successive magnetograms,
\cite{Demoulin2003} argued that the apparent motion of magnetic flux
from one magnetogram to the next over $\Delta t$ could be described by
a ``footpoint velocity'' $\uvec$, that is related to the plasma
velocity $\vvec$ by
\be \uvec B_n = \vvec B_n - v_n \bvec_h ~, \label{eqn:uvec} \ee 
where $\hatn$ denotes the unit vector normal to the surface, the $n$
subscript denotes the normal component of a vector, and the $h$
subscript denotes the horizontal components of a vector, which are
perpendicular to $\hatn$.  This assumption can be
used to express the normal component of the magnetic induction
equation in terms of a continuity-like, finite-difference equation,
\be \frac{\Delta B_n}{\Delta t} + \nabla_h \cdot (\uvec B_n) = 0
~, \label{eqn:ctty} \ee 
where $\Delta B_n = B_n(t_f) - B_n(t_i)$ is the change in the normal
magnetic field between the initial time $t_i$ and the final time
$t_f$, and $\nabla_h$ is approximated with finite differences in the
spatial coordinates orthogonal to the normal direction.  The analogy
with the continuity equation is not precise, because unlike number or
mass densities, which are always positive, magnetic flux can be
positive or negative.  So, unlike total particle number or total mass
in a volume, neither total signed nor total unsigned flux is conserved
in a magnetogram: existing opposite polarity fluxes can cancel, or new
bipolar flux can emerge.  

\citet{Demoulin2003} argued that tracking methods applied to
magnetograms estimate $\uvec$, instead of $\vvec_h$ (cf.,
\citealt{Chae2001}), but \citet{Schuck2005, Schuck2008} argued that
some tracking methods return a biased estimate, $\tilde \uvec$, with
the contribution from $\vvec_h$ weighted more heavily than in the true
$\uvec.$ The term ``footpoint velocity'' implicitly refers to
extensions of magnetic flux above the magnetogram surface.
\citet{Welsch2006} used the alternative ``flux transport velocity'' to
refer to $\uvec$ without reference to field structure above or below
the magnetogram.  For purposes of tracking, we assume the LOS field,
$B_\ell$, derived from the NFI observations, approximates the normal
field $B_n$.

We used the FLCT code \citep{Welsch2004, Fisher2008} to estimate flux
transport velocities for this dataset.  The FLCT code uses Fourier
cross-correlation of $B_n(t_i)$ with $B_n(t_f)$ near a given pixel,
say one centered at $(x_j, y_k)$, to determine the two-component
displacement $\Delta \svec$ of magnetic field structure between the
magnetograms near $(x_j, y_k)$; division by $\Delta t$ then results in
a velocity estimate at $(x_j, y_k)$.  The correlation is localized by
first weighting each magnetogram by a Gaussian windowing function,
$\exp(-r_{jk}^2/\sigma^2)$, where $r_{jk}$ is distance from $(x_j,
y_k)$, and $\sigma$ is a free parameter.  Optionally, low-pass
filtering, with a Gaussian roll-off, can be applied to the Fourier
spectra prior to performing the cross-correlation.  In tests applying
shifts to quiet-Sun magnetograms, \citet{Fisher2008} found roll-off
values of $0.2 k_{\rm max}$ -- $0.5 k_{\rm max}$ improved recovery of
the applied shifts, where $k_{\rm max}$ is the highest wavenumber each
dimension.  In this work, we used a roll-off wavenumber of 0.5 $k_{\rm
  max}$, which corresponds to relatively weak Fourier filtering.

Like FLCT, other tracking methods typically apply an apodization or
windowing function, the spatial extent of which is a free parameter,
to the initial and final magnetograms, and find a velocity that
extremizes some functional over this window \citep{Schuck2006}.  Given
data with sufficient cadence, the interval $\Delta t$ between pairs of
magnetograms to be tracked can also be varied.  Tracking methods
therefore estimate velocities from the effects of actual velocities
averaged both spatially over the windowed region and temporally over
the time interval $\Delta t$ between magnetograms.  This averaging
implies that flows on spatial scales smaller than the windowing
function, and with lifetimes shorter than $\Delta t$, cannot be
recovered in the flow estimation process.

Beyond these algorithmic aspects of the tracking process, it can be
also seen that noise and other artifacts in magnetogram measurements
will contribute to the difference $\Delta B_n$ in equation
(\ref{eqn:ctty}), implying such effects can influence estimated
velocities.

These considerations motivate efforts to understand the interplay
between windowing and tracking cadence and flow spatial scale and
lifetime, as well as the effects of magnetogram noise.  To investigate
these relationships, we estimated flux transport velocities from pairs
of magnetograms that were separated in time by eight varying
intervals, with $\Delta t \in \{$2, 4, 8, 16, 32, 64, 128, 256$\}$
minutes, that were both unbinned and binned to six varying macropixel
sizes $\Delta x \in \{$2, 4, 8, 16, 32, 64 $\}$ in terms of the data's
original $0.16 \arcsec$ pixels, with four choices of apodization
parameter $\sigma \in \{$2, 4, 8, 16 $\}$.  Pixel values were simply
averaged when rebinned, which does not exactly mimic how magnetographs
with either worse spatial resolution or poor weak-field sensitivity
would detect magnetic flux.

Since we are interested in characterizing the effects of noise on
estimated flows, we tracked all pixels above a threshold of 15 Mx
cm$^{-2}$, which is both near our estimate of the noise level, and
includes the spurious weak-field regions in the sunspot cores.

\section{Lifetimes of Magnetic Structures}
\label{sec:b_lifetimes}

\begin{figure}[ht]
  \centerline{\psfig{figure=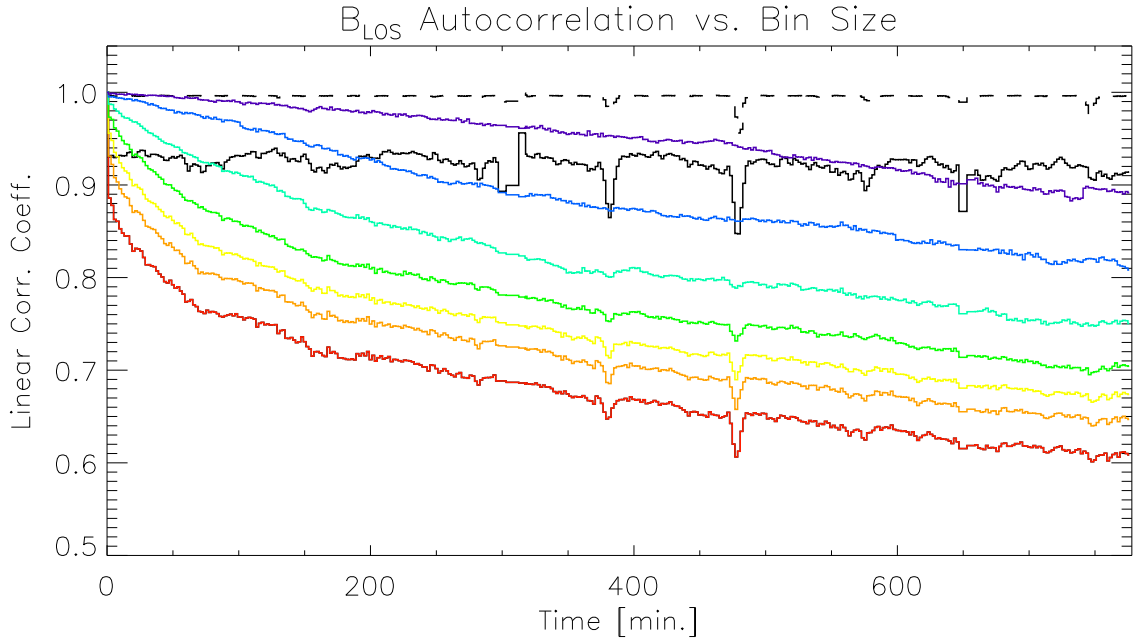,width=6.5in}}
  \caption[]{\footnotesize \textsl{Colored curves show rank-order
      autocorrelation coefficients with increasing lag times, for
      magnetograms at several spatial binnings: red is for $0.16
      \arcsec$-pixel data; orange, yellow, green, aqua, blue, and
      purple are for data rebinned $2 \times 2, 4 \times 4, 8 \times
      8, 16 \times 16, 32 \times 32,$ and $64 \times 64$,
      respectively.  The black dashed and solid lines show
      frame-to-frame linear and rank-order correlation coefficients,
      respectively, for data binned $2 \times 2$. Note the limited
      range of the vertical axis. Downward spikes before $500$ minutes
      arise from frames with many cosmic ray hits. The effects of the
      X-class flare can be seen as small blips near $743$ minutes.}}
        \label{fig:mag_auto}
\end{figure}

To provide context for the time evolution of flows on varying spatial
scales, we autocorrelated the magnetograms themselves, at several
spatial resolutions: with $0.16 \arcsec$ pixels, and then with data
rebinned to macropixels $ \Delta x \in \{ 2, 4, 8, 16, 32, 64 \}$
pixels on a side.  While each magnetogram is a 2D array, we computed
the correlations as if both were 1D arrays.  All pixels outside of the
bubble region were used, not just those above the noise threshold used
for tracking.  The colored curves in Figure \ref{fig:mag_auto} show
rank-order correlation coefficients for the autocorrelations at each
spatial resolution for increasing lag time.  In addition, the black
dashed and solid curves show frame-to-frame linear and rank-order
correlations, respectively, for the $2 \times 2$ binned data. Note the
limited range of the $y$-axis.

The high frame-to-frame correlations imply that the magnetic field
does not evolve much from one magnetogram to the next.  It can also be
seen that frame-to-frame linear correlations hover around unity, but
frame-to-frame rank-order correlations are significantly lower,
implying rank-order correlation is a more sensitive test of
differences between magnetograms.  This plot also shows that the
lifetime of this active region's magnetic structure, over the entire
magnetogram field of view, is much longer than the duration of the
dataset that we analyze.

Several artifacts are also present in these time series.  Downward
spikes in correlations correspond to frames in which isolated pixels
contained large fluctuations, probably due to hits by cosmic rays.
Despiking the magnetograms can remove these spikes, but here we are
interested in the effects of such artifacts on autocorrelations.
Larger binnings average out these effects. In terms of lag
times plotted here, the start time of X-class flare that began around
02:28UT on 2006-Dec-13 corresponds to $743$ minutes, and its
effects can be seen as small wiggles in some autocorrelation curves
near the right end of the plot.

\begin{figure}[ht]
  \centerline{\psfig{figure=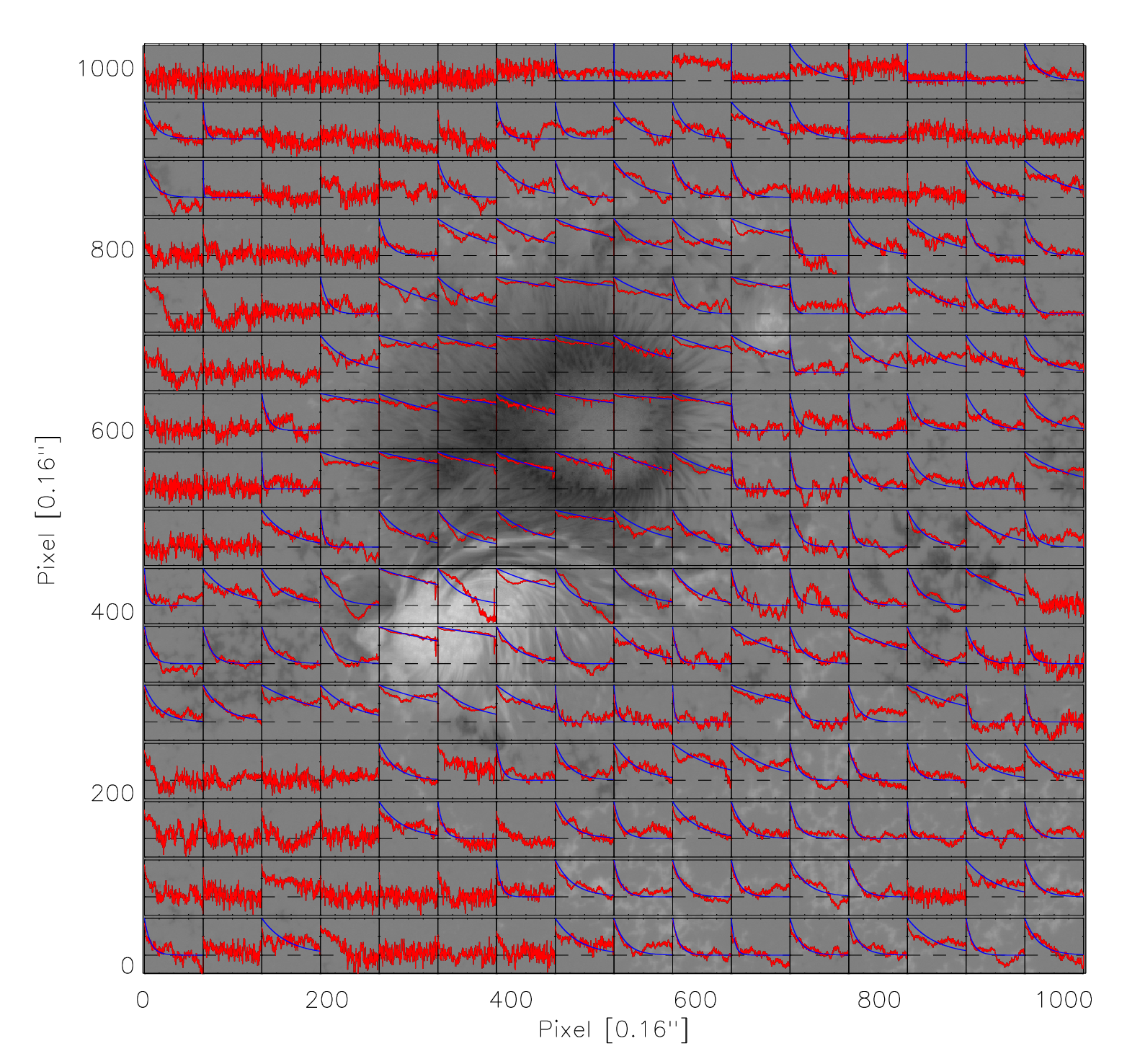,width=5.5in}}
  \caption[]{\footnotesize \textsl{The grayscale image is $|B_{\rm
        LOS}|$ in a full-resolution magnetogram, with saturation at
      $\pm 4$ kG.  Red curves show rank-order autocorrelations of
      $B_{\rm LOS}$, in $(32 \times 32)$-binned subregions of ($2
      \times 2$)- binned data, vs. lag time, up to the duration of the
      dataset. The vertical range in each cell is [-0.5, 1.0],
      with a dashed black line at zero correlation. Blue curves show
      one-parameter fits to the decorrelation in each subregion,
      assuming exponential decay, with the decay constant as the only
      free parameter.  Only subregions with median occupancy of at
      least 20\% of pixels above our 15 G threshold were fit. As
      expected, field structures persist longer in stronger-field
      regions.}}
        \label{fig:mag_256}
\end{figure}
\begin{figure}[ht]
  \centerline{\psfig{figure=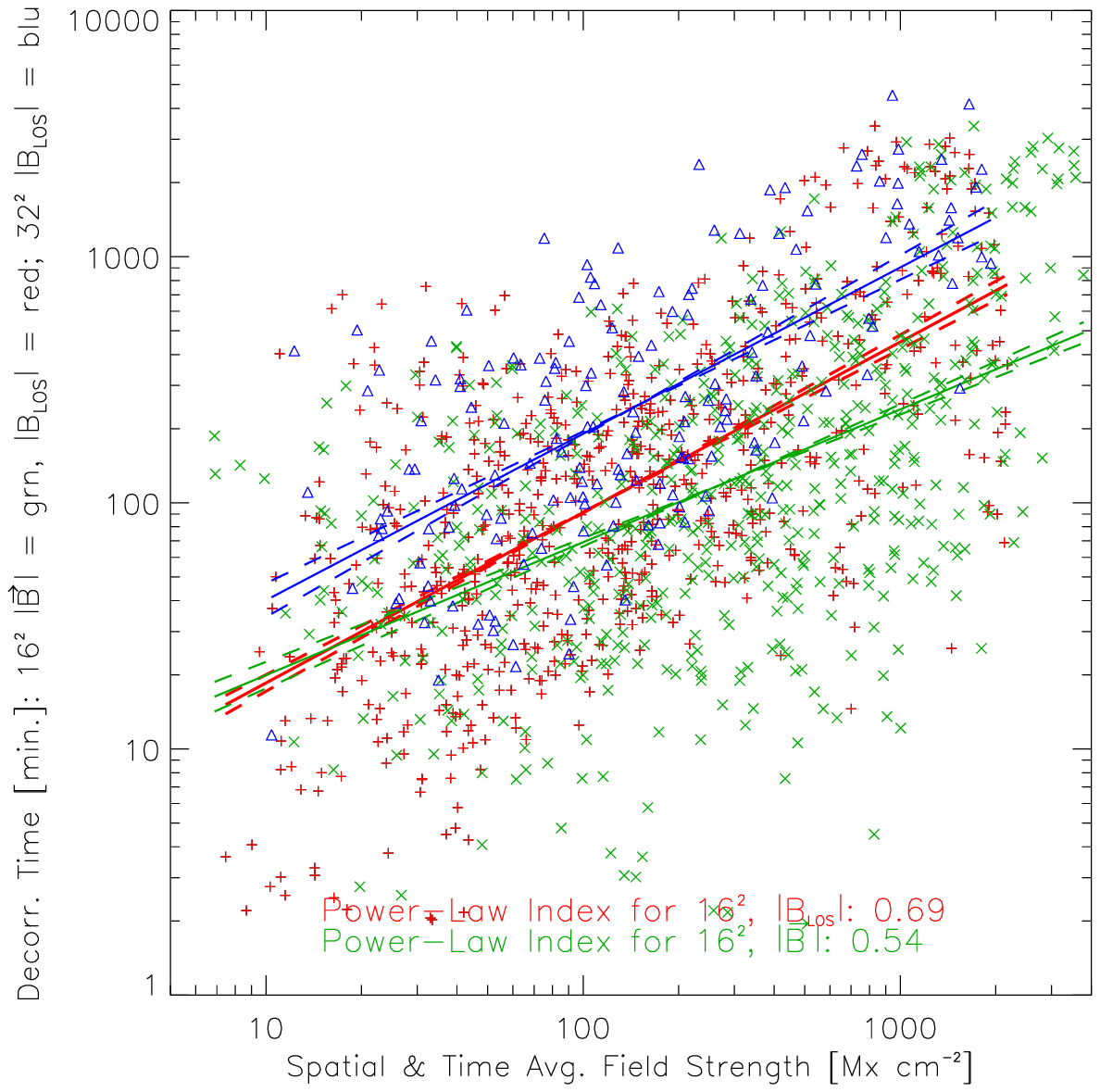,width=4.5in}}
  \caption[]{\footnotesize \textsl{Fitted lifetimes of boxcar-averaged
      $|B_{\rm LOS}|$ and $|\bvec|$, averaged over each subregion, for
      the non-bubble region of the FOV.  Blue triangles correspond to
      $(32 \times 32)$-binned subregions (from fits in Figure
      \ref{fig:mag_256}) of $|B_{\rm LOS}|$. Red +'s correspond to
      $(16 \times 16)$-binned subregions $|B_{\rm LOS}|$
      (autocorrelations and fits at this resolution are not plotted in
      this paper), and green $\times$'s correspond to $|\bvec|$
      averaged over $(16 \times 16)2$-binned subregions.  Fitted
      exponents to log-log values of $|B_{\rm LOS}|$ and $|\bvec|$ for
      $(16 \times 16)$ bins are shown with corresponding
      colors. Dashed lines correspond to uncertainties in each fitted
      slope, based upon $\chi^2$; the slope of the blue fit (0.68) is
      similar to the red fit (0.69).}}
        \label{fig:mag_life_vs_b}
\end{figure}

Distinct types of magnetic regions --- notably, sunspots, plage, and
quiet Sun --- all influence the autocorrelations in Figure
\ref{fig:mag_auto}. By investigating the lifetimes of magnetic
structures on smaller spatial scales, we can better understand the
lifetimes of magnetic structures in each type of region.  Accordingly,
in Figure \ref{fig:mag_256}, we show autocorrelations (red curves) for
$B_{\rm LOS}$ in $(32 \times 32)$-binned subregions of ($2 \times
2$)-binned data (so each subregion is $(64 \times 64)$
full-resolution pixels).  In the same figure, we also show
one-parameter fits (blue curves) to the decorrelation in each
subregion, assuming the autocorrelation function decays exponentially
with lag time $\Delta T$, $\exp(-\Delta T/\tau)$, with the decay
constant $\tau$ as the only free parameter.  In each subregion, we
take the fit decay constant, $\tau$, to be the lifetime of the
structure in that subregion.  We define the occupancy of a subregion
to be the percentage of pixels above our 15 G noise threshold.  Only
pixels with a median occupancy, over the dataset's duration, of at
least 20\% were fit.  Background gray contours in the figure show 50 G
and 200 G levels of $|B_{\rm LOS}|$ from the initial magnetogram (in
full-resolution pixels), so variations in lifetimes can be understood
in the context of active region structure.

We then investigated the dependence of magnetic lifetime on field
strength.  To start, we computed a measure of subregion field strength
by boxcar averaging flux densities from $|B_{\rm LOS}|$ from the five
initial magnetograms in our sequence at $(2 \times 2)$ resolution,
then taking the absolute value, then computing the spatial average of
this unsigned, boxcar-averaged flux density over each subregion.  We
chose to use this measure of initial field strength to address the
question: ``For a given current field strength, how long will a
structure persist?''  The relatively short fitted lifetimes of some
magnetic regions suggests that averaging over longer intervals ---
e.g., the length of the entire run --- would imply averaging over
several magnetic turnover times.  It can be seen in Figure
\ref{fig:mag_256} that some of the fits represent the evolution
poorly, but many fits match the observations well, given the single
free parameter.  Of the 195 subregions with at least 20\%
above-threshold occupancy, only four have fitted lifetimes that are
less than the two-minute nominal cadence (see, e.g., the two
subregions to the left of the top-right subregion).  We ignore these
too-short fits as pathological.  In Figure \ref{fig:mag_life_vs_b}, we
plot the fitted lifetimes as a function of these spatially and
temporally averaged subregion fields (blue +'s).  A linear fit between
the logarithms of each returns a power-law exponent of 0.66.  The solid
blue line shows the fit, while the dashed blue lines reflect
uncertainty in the fitted slope, based upon $\chi^2$ computed from
the fit.
This trend is consistent with the idea that convection
disrupts magnetic field structures, but stronger magnetic fields more
strongly inhibit convection \citep{Title1992, Berger1998, Welsch2009},
so structures persist longer in stronger-field regions.
Despite the trend of increasing lifetime with field strength, we note
some high-field-strength regions have short lifetimes.  Relatively
rapid evolution in these high-field strength regions might play some
role in flare activity, but pursuing this question is beyond the scope
of this paper.

To investigate the dependence of this field-strength-versus-lifetime
relationship on spatial scale, we repeated this procedure with the
subregion size reduced to $16^2$ ($2 \times 2$)-binned pixels
(autocorrelations and fits at this resolution are not shown).
Lifetimes as a function of subregion-averaged field strength are
plotted in red in Figure \ref{fig:mag_life_vs_b}, excluding 28
subregions with fitted lifetimes $< 2$ min (of 733 subregions with at
least 20\% above-threshold occupancy). A linear fit between the logarithms
of field strengths and lifetimes returns a power-law exponent of 0.69
in this case. The solid red line shows the fit, while the dashed lines
reflect uncertainty in the fitted slope.  The trend of lifetime versus
field strength did not change much for subregions smaller by a factor
of two.

Recalling that our estimates of $|B_{\rm LOS}|$ suffer from
non-monotonicity in the Stokes $V/I$ ratio, we also characterized the
lifetimes of NFI structures as functions of $|B_z|$ and $|\bvec|$ from
the SP data, in subregions averaged to the same spatial resolution
($16^2$ regions of $2 \times 2$ pixels).  Note, however, that the SP
data were recorded near the middle of our observing run (between 20:30
and 21:14 UT), while the magnetic lifetimes were computed from the
beginning of our run.  Also, our treatment takes no account of spatial
filling factors of magnetic structures within resolution elements.
Despite these uncertainties in the SP field strengths, fits of
lifetimes as a function of $|B_z|$ (not shown) were similar to those
for $|B_{\rm LOS}|$.  The solid green line in Figure
\ref{fig:mag_life_vs_b} shows the fitted slope (dashed lines reflect
uncertainties in the fitted slope) of lifetimes as a function of
$|\bvec|$.  The slightly weaker dependence of lifetime on $|\bvec|$
compared to $|B_{\rm LOS}|$ might arise from either: vertical magnetic
fields inhibiting convection more efficiently than horizontal magnetic
fields; or rapid magnetic evolution near polarity inversion lines
(PILs) of the normal field, regions where the magnetic field is
predominantly horizontal and where flux emerges or submerges; or both.
We note that the range of $|\bvec|$ is greater than for either of the
other two quantities, and this longer run will naturally decrease the
fitted slope for the same rise. \cite{Demoulin2003} suggested that
in regions with predominantly horizontal magnetic fields, apparent
motions of vertical magnetic flux could be rapid.

We remark that autocorrelation analyses from observed magnetograms
like those presented here could be compared to autocorrelations from
synthetic magnetograms extracted from simulations of magnetoconvection
(e.g., \citealt{Rempel2011b}), as a test of statistical consistency
between the observations and simulations.

\section{Flow Lifetimes}
\label{sec:flow_lifetimes}


In this section, we first consider selection of tracking parameters,
and then investigate lifetimes of estimated flows versus spatial scale
and magnetic field strength.

\subsection{Selection of Tracking Parameters}
\label{subsec:params}

\citet{Welsch2004} noted that flows derived by tracking alone only
approximately solve equation (\ref{eqn:ctty}). \cite{Schuck2006}
suggested that, in the presence of magnetogram noise, exactly solving
the induction equation is probably not optimal. \citet{Welsch2007}
suggested that for fixed $\Delta t$ (i.e., imposed by available data),
$\sigma$ could be chosen to optimize agreement between the observed
$\Delta B_n/\Delta t$ and the estimated $\nabla_h \cdot (\uvec B_n)$.
When observations are frequent enough that $\Delta t$ can also be
varied, considerations beyond optimizing consistency with equation
(\ref{eqn:ctty}) are useful.  

In addition, \citet{Welsch2011} demonstrated that enforcing strict
consistency with equation (\ref{eqn:ctty}) in deriving flows is
probably unwise.  Using 96-minute cadence, full-disk magnetogram
sequences observed by MDI \citep{Scherrer1995}, they characterized
magnetic evolution in three active regions near disk center.  In
addition to using FLCT to estimate flows, they also numerically
derived ideal electric fields to exactly solve Faraday's law, from a
Poisson equation, with a source given by their estimate of the radial
magnetic field evolution $[\Delta B_R/\Delta t]$, for the inductive
part of the electric field \citep{Welsch2007}.  Where
\citet{Welsch2009} found flow fields derived by either FLCT or DAVE
from 96-minute cadence magnetograms to be correlated from one
96-minute interval to the next, the ``Poisson flow fields'' that
\citet{Welsch2011} derived to be exactly consistent with $\Delta
B_R/\Delta t$ decorrelated from one frame to the next.  
Assuming random fluctuations (due to, e.g., noise present in the estimated
flows), it can be shown that, if evolution in the flow is negligible
(so the only difference between the flow maps is due to random errors
in the estimated flows), then even a relatively high correlation
coefficient (e.g., 0.9) implies significant fluctuations present
(respectively, 33\% of the variance in the flow component).  It can
also be shown that even allowing for evolution in the flow only
moderately reduces the fraction of decorrelation that must be ascribed
to random fluctuations in the flow estimates.
The fact that FLCT and DAVE flows did not decorrelate between frames,
but the Poisson flows did, implies the latter contained significant
random fluctuations, and were therefore probably strongly influenced
by magnetogram noise.  This suggests that using equation
(\ref{eqn:ctty}) as the sole guide to determining tracking parameters
has limitations.

We therefore consider the effect of the time interval $\Delta t$
between each pair of tracked magnetograms on the relationship between
flux transport velocities $\uvec$ from tracking and plasma velocities
$\vvec$ in equation (\ref{eqn:uvec}).  Essentially, if $\Delta t$ is
either too short or too long, then equation (\ref{eqn:uvec}) is
invalid.  To explain why, we consider each case separately.

When the time difference $\Delta t$ between magnetograms is too short,
very little magnetic evolution has occurred (as in the black curves in
Fig. \ref{fig:mag_auto}), so most of the difference $\Delta B_n$
between initial and final magnetograms that appears in equation
(\ref{eqn:ctty}) is due only to noise.  Consequently, flux transport
velocities inferred from $\Delta B_n/\Delta t$ will be spurious.  For
instance, a transient spike in the field strength of a single pixel in
a weak unipolar region would lead to inference of a high-speed
converging flow surrounding that pixel, consistent with the sudden
field increase, followed by a diverging flow, consistent with the
field decrease after the spike disappears.  In general, these
noise-induced flows will be unrelated to actual plasma velocities.
Consequently, if $\Delta t$ is too short, then noise is a large
component of magnetogram difference $\Delta B_n$, and we refer to
flows estimated in this regime as noise-dominated.

When the time difference $\Delta t$ between magnetograms is too long,
displacements of magnetic flux might exceed a pixel length, meaning
the finite difference approximation used in equation (\ref{eqn:ctty})
could be inaccurate.  In this regime, which we refer to as
displacement-dominated, flux transport velocities estimated by
tracking are more properly described as time averages of true flux
transport velocities, which acted and evolved over timescales shorter
than $\Delta t$.  It should be noted that, although estimated flows in
the displacement-dominated regime do not correspond directly to plasma
velocities, they still have a clear physical meaning: they represent
the time-averaged flow over timescales of a flow lifetime or longer.
We expect that motions derived for $\Delta t$ longer than a flow
lifetime will underestimate true flow speeds: the true path of a
feature executing zig-zag motion can be resolved with short enough
$\Delta t$, but if $\Delta t$ is too long to resolve each step on the
path, only the (shorter) net displacement is recovered (cf., the full
path length traversed over that time interval).  Hence, a given $\Delta
t$ can average over the effects of shorter-lived flows.
\begin{figure}[ht]
  \centerline{\psfig{figure=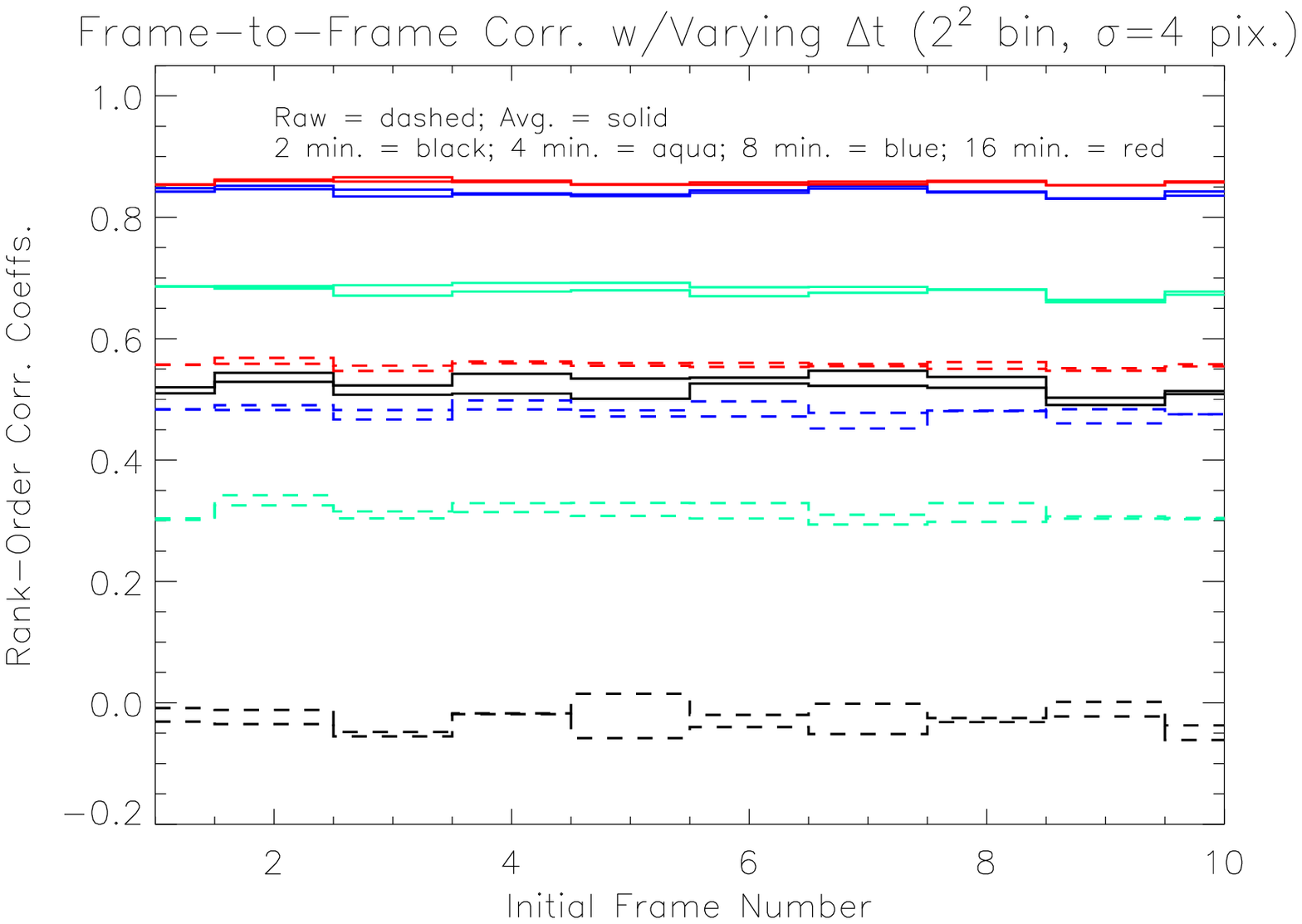,width=5.0in}}
  \caption[]{\footnotesize \textsl{Plotted lines show frame-to-frame
      correlation coefficients for the $x$- and $y$- components of
      estimated flows, for varying time intervals $\Delta t$ between
      tracked magnetograms --- 2 min in black, 4 min in aqua, 8 min in
      blue, 16 min in red.  Dashed curves are from flows derived from
      un-averaged magnetograms, while solid curves are from flows
      derived from boxcar-averaged magnetograms.  Flow estimates are
      more stable with averaged magnetograms and with longer $\Delta
      t$, until saturation occurs around $\Delta t$ = 8 min.}}
        \label{fig:hicad_avg}
\end{figure}

It would be convenient if an ideal choice of $\Delta t$ could be made
{\em a priori}. As previously noted, however, flows with a range of
speeds, and time and length scales are present at the solar
photosphere (e.g., granular flows with speeds $\sim 1$ km sec$^{-1}$
with lifetimes of a few minutes and $\sim 1$ Mm length scales, and
supergranular flows with speeds a few times $10^2$ m sec$^{-1}$ with
lifetimes of about a day, and length scales of $\sim$ 10 Mm).  Poor
choice of $\Delta t$ can, however, be determined by {\em post facto}
analysis of tracking results, meaning trial runs can be used to
identify suitable timescales.

\begin{figure}[ht]
  \centerline{\psfig{figure=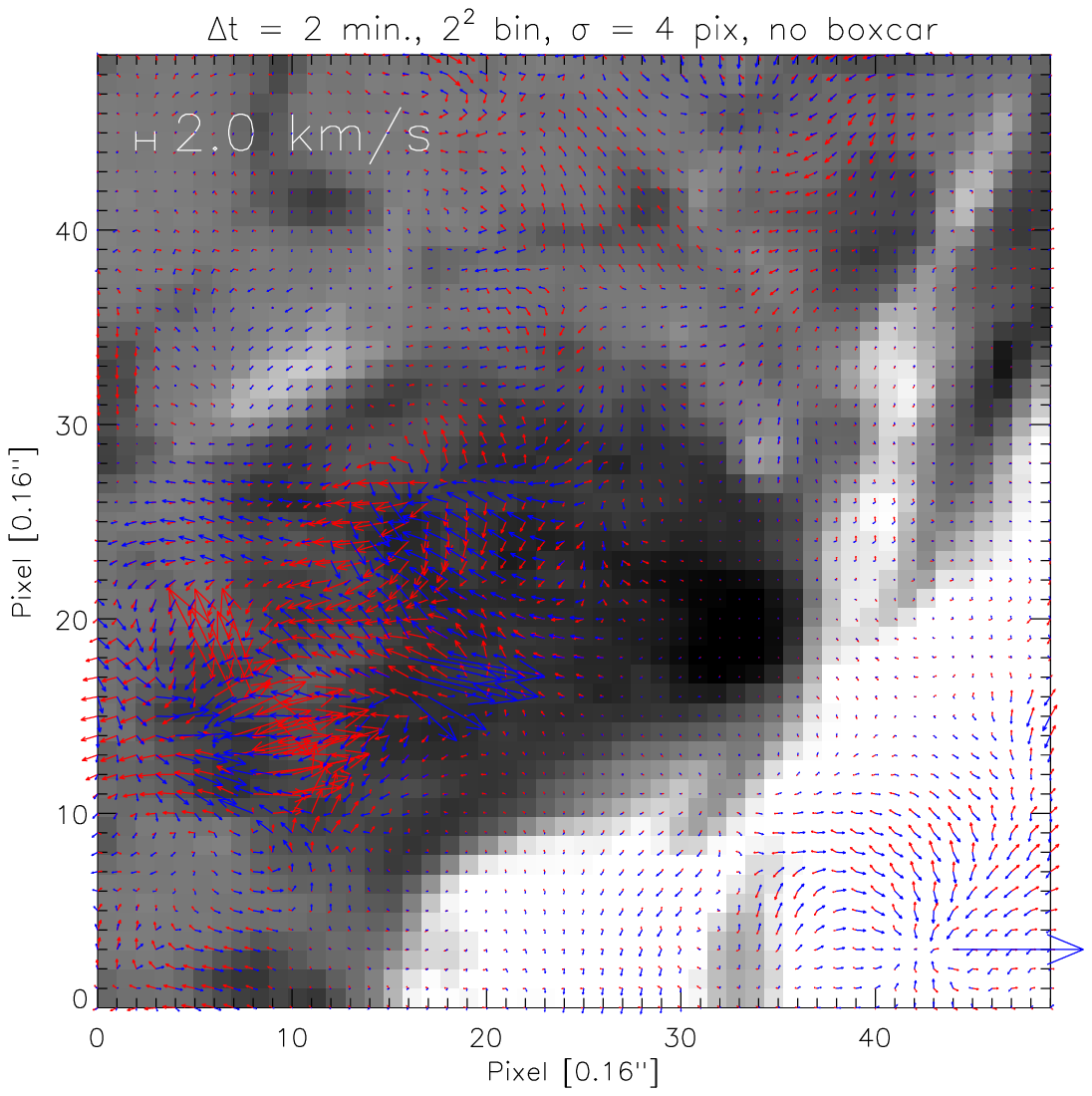,width=5.0in}}
  \caption[]{\footnotesize \textsl{Successive flow maps in
      a subregion of the full field of view, derived from unaveraged
      magnetograms (binned $2 \times 2$) separated in time by $\Delta
      t = 2$ min with $\sigma = 4$ pixels.  Initial flows are shown
      by red vectors, subsequent flows are shown by blue vectors.
      Note significant variations between the flow fields in this
      noise-dominated regime.}}
        \label{fig:noise_dom}
\end{figure}

\begin{figure}[ht]
  \centerline{\psfig{figure=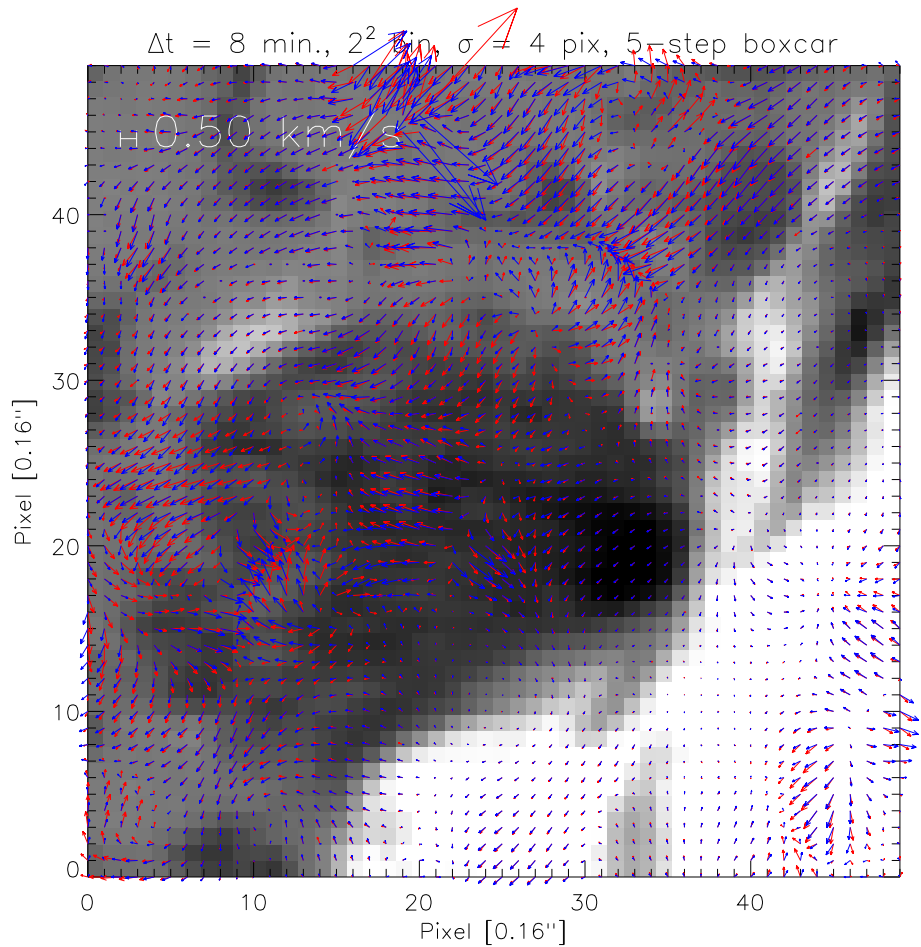,width=5.0in}}
  \caption[]{\footnotesize \textsl{Successive flow maps from the same
      subregion, derived from averaged magnetograms (binned $2 \times
      2$) separated in time by $\Delta t = 8$ min with $\sigma = 4$
      pixels.  Initial flows are shown by red vectors, subsequent
      flows are shown by blue vectors.  Note significant agreement
      between flow fields.}}
        \label{fig:noise_ok}
\end{figure}

We first consider the noise-dominated regime, in which correlations
between each velocity component from successive flow estimates will be
small.  Frame-to-frame correlations can be improved by increasing
$\Delta t = t_f - t_i$, and by averaging successive magnetograms
around $t_i$ and $t_f$ to synthesize initial and final images.  Figure
\ref{fig:hicad_avg} shows frame-to-frame, rank-order correlations for
the $x$- and $y$-components of estimated velocities, with $\sigma$
fixed at 4 pixels but for several $\Delta t$'s, using either
unaveraged (dashed) and averaged (solid) magnetograms for $B(t_i)$ and
$B(t_f)$.  Low correlations for short values of $\Delta t$ imply that
noise has degraded the velocity estimates.  For $\Delta t = 2$ min,
the flow fields are even weakly anticorrelated.  Increasing $\Delta t$
clearly increases frame-to-frame correlations, because real magnetic
field evolution becomes more significant than spurious evolution from
noise.  Saturation occurs, however: doubling $\Delta t$ from 8 to 16
minutes did not greatly improve frame-to-frame correlations.  As
$\Delta t$ increases further, into the displacement-dominated regime,
frame-to-frame correlations can approach unity, since displacements
averaged over very long time intervals do not change much if the
intervals' endpoints are shifted slightly.  To average the
magnetograms, a 3-magnetogram boxcar was used for $\Delta t$ = 2 and 4
min, while a 5-magnetogram boxcar was used for $\Delta t$ = 8 and 16
min.  For cadences of 2, 4, and 8 minutes, there is overlap in data
used for $B(t_i)$ and $B(t_f)$.  This probably leads to blurring of
features from their motion over the summing interval, and could also
lead to correlated noise in $B(t_i)$ and $B(t_f)$ \citep{Schuck2006}.
The improvement in frame-to-frame correlations from averaging
magnetograms is substantial.  Linear correlation coefficients (not
shown) were both lower and less consistent across frames.  We
therefore employ rank-order correlations to compute all subsequent
autocorrelations.  We provide concrete examples of these effects in
Figures \ref{fig:noise_dom} and \ref{fig:noise_ok}.  In the former,
successive flow fields computed with $\Delta t = 2$ min show very
little consistency between flow fields.  In the latter, successive
flows computed with $\Delta t = 8$ min are much more consistent.
Generally, frame-to-frame correlations are higher with larger values
of $\sigma$ (not shown).  This is consistent with FLCT detecting flows
on larger spatial scales with larger $\sigma$, and larger-scale flows
persisting longer in time.  We discuss this issue in greater detail below.

\begin{figure}[ht]
  \centerline{\psfig{figure=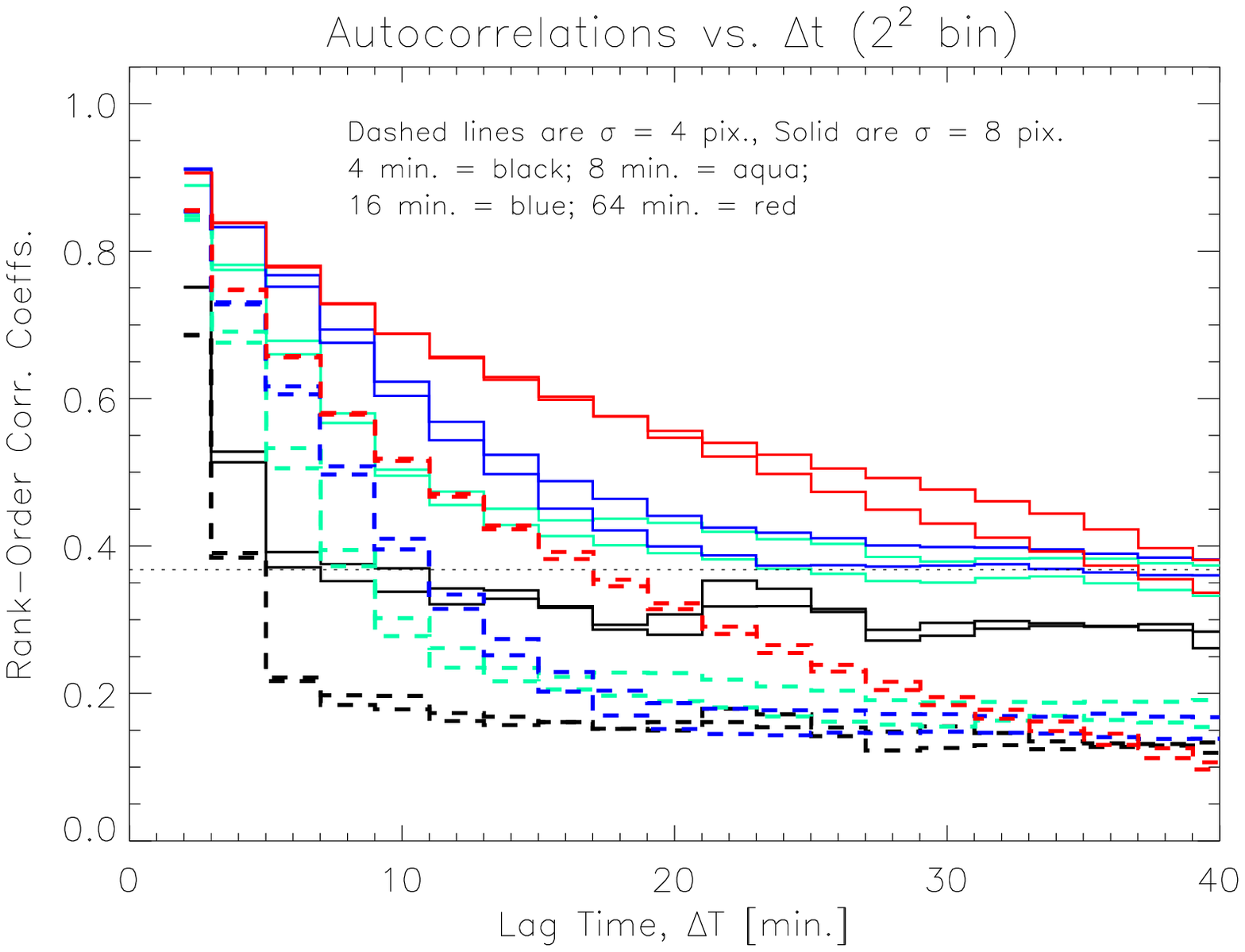,width=5.0in}}
  \caption[]{\footnotesize \textsl{Plotted lines show correlation
      coefficients for the $x$- and $y$- components of flows estimated
      at $t_i$ and $t_i + \Delta T$, for several time intervals
      $\Delta t$ between tracked magnetograms --- 4 min in black, 8
      min in aqua, 16 min in blue, 64 min in red.  Dashed curves
      are from flows derived with $\sigma = 4$ pixels, while solid
      curves are from flows with $\sigma = 8$ pixels.  All flows were
      derived from boxcar-averaged magnetograms.  Horizontal dotted
      line is $1/e$.}}
        \label{fig:bigdt_autocorr}
\end{figure}

In the displacement-dominated regime, correlations between each
velocity component from flow fields separated by lag times $\Delta T$
much shorter than $\Delta t$ can be high, but such correlations will
be very weak when $\Delta T$ is on the order of $\Delta t$ or longer.
This implies substantial flow evolution has occurred over the $\Delta
t$ between magnetograms.  Phrased another way, $\Delta t$ is longer
than the estimated flows' lifetimes.  For estimated flows to
reflect actual plasma velocities, however, $\Delta t$ cannot exceed the
flow lifetime.  Displacement-dominated flows are still physically
meaningful, however: they quantify the average effects of flows over
the course of more than one flow lifetime.

Having demonstrated the advantages of averaging, we use flows derived
from magnetograms averaged with a five-step boxcar in our analyses of
flow lifetimes below. For convenience, we will sometimes refer to
flows inferred by FLCT as velocities, but the reader should keep in
mind that FLCT's flows should not necessarily be interpreted as
estimates of plasma velocities at the center of the time interval
$\Delta t$.

\subsection{Estimating Flow Lifetimes}
\label{subsec:lifetimes}

We define the lifetime $\tau$ of a flow as the lag time $\Delta T$ at
which a one-parameter fit to the autocorrelation drops to $1/e \simeq
0.37$.   In Figure \ref{fig:bigdt_autocorr}, we plot autocorrelations
for flow estimates made with two choices of $\sigma$ and several
choices of $\Delta t$.  For $\Delta t = 64$ minutes, the lifetimes of
the $x$- and $y$- flow components are around 35 minutes for $\sigma =
8$, implying that flow estimates made with $\Delta t = 64$ minutes are
displacement-dominated.  In contrast, the lifetimes of $\sigma=8$
flows estimated with $\Delta t =$ 4, 8, and 16 minutes are all at
least as long as $\Delta t$.  Lifetimes of flows estimated with
$\sigma = 4$ (dashed lines) are about half as long as $\sigma = 8$
estimates for $\Delta t =$ 8, 16, and 64 minutes.

\begin{figure}[ht]
  \centerline{\psfig{figure=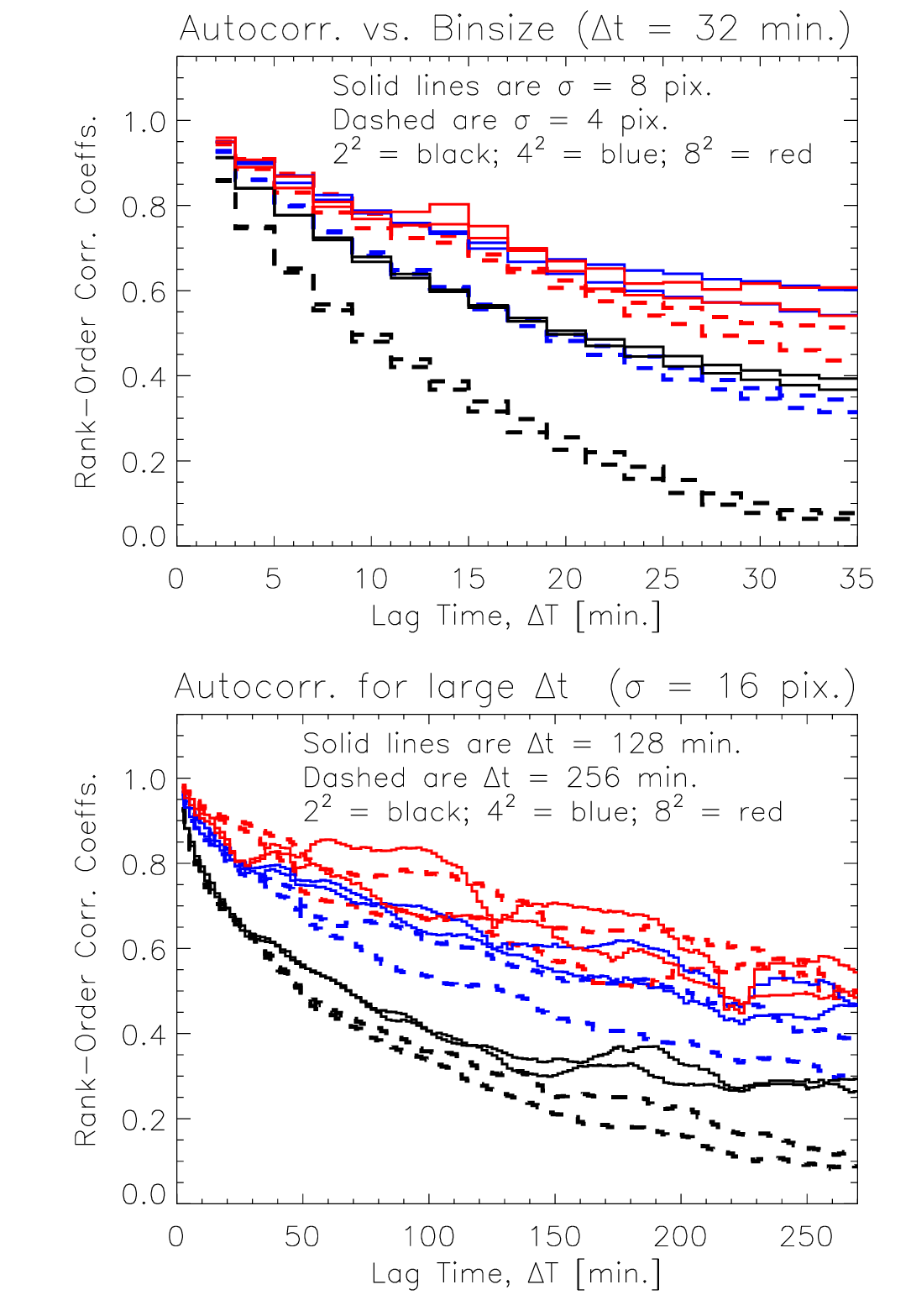,width=5.5in}}
  \caption[]{\footnotesize \textsl{Top: Autocorrelations of $x$- and
      $y$-components of flows derived using two values of $\sigma$
      (dashed = 4 pix., solid = 8 pix.) from magnetograms binned into
      different-sized macropixels (black = $2^2$, blue = $4^2$, red =
      $8^2$).  Evidently, keeping $\sigma$ fixed but doubling bin size
      approximates keeping bin size fixed but doubling $\sigma$.
      Additional binning did not improve correlations.  Bottom:
      Autocorrelations of $x$- and $y$-components of flows estimated
      with large values of $\sigma$ (16 pix. for all flows here) and
      from data rebinned into larger macropixel sizes (colors as in
      top panel) for $\Delta t = 128$ min (solid) and $\Delta t =
      256$ min (dashed) are long-lived.  As in top panel, additional
      binning did not improve correlations. All flows in both panels
      were derived from boxcar-averaged magnetograms.}}
        \label{fig:binsigma}
\end{figure}

The autocorrelations in Figure \ref{fig:bigdt_autocorr} also suggest
that flows estimated on larger spatial scales --- using a larger value
of $\sigma$ --- have longer lifetimes.  In fact, this is typical.  In
addition, longer flow lifetimes are also found when data are rebinned
into larger macropixels, though there is a point at which rebinning
does not increase decorrelation times.  The top panel of Figure
\ref{fig:binsigma} demonstrates that choosing larger values of
$\sigma$ or rebinning into larger macropixels are practically
equivalent.  The bottom panel of Figure \ref{fig:binsigma} implies
that, when flows are displacement-dominated (i.e., they decorrelate at
lag times $\Delta T < \Delta t$), longer-lived flows can be identified
in rebinned data.  We note that increasing the windowing parameter
$\sigma$ can increase compute time for FLCT (and probably other
tracking methods), while rebinning (holding $\sigma$ fixed) can
decrease compute time, since fewer velocity estimates must be made.

\begin{figure}[ht]
  \centerline{\psfig{figure=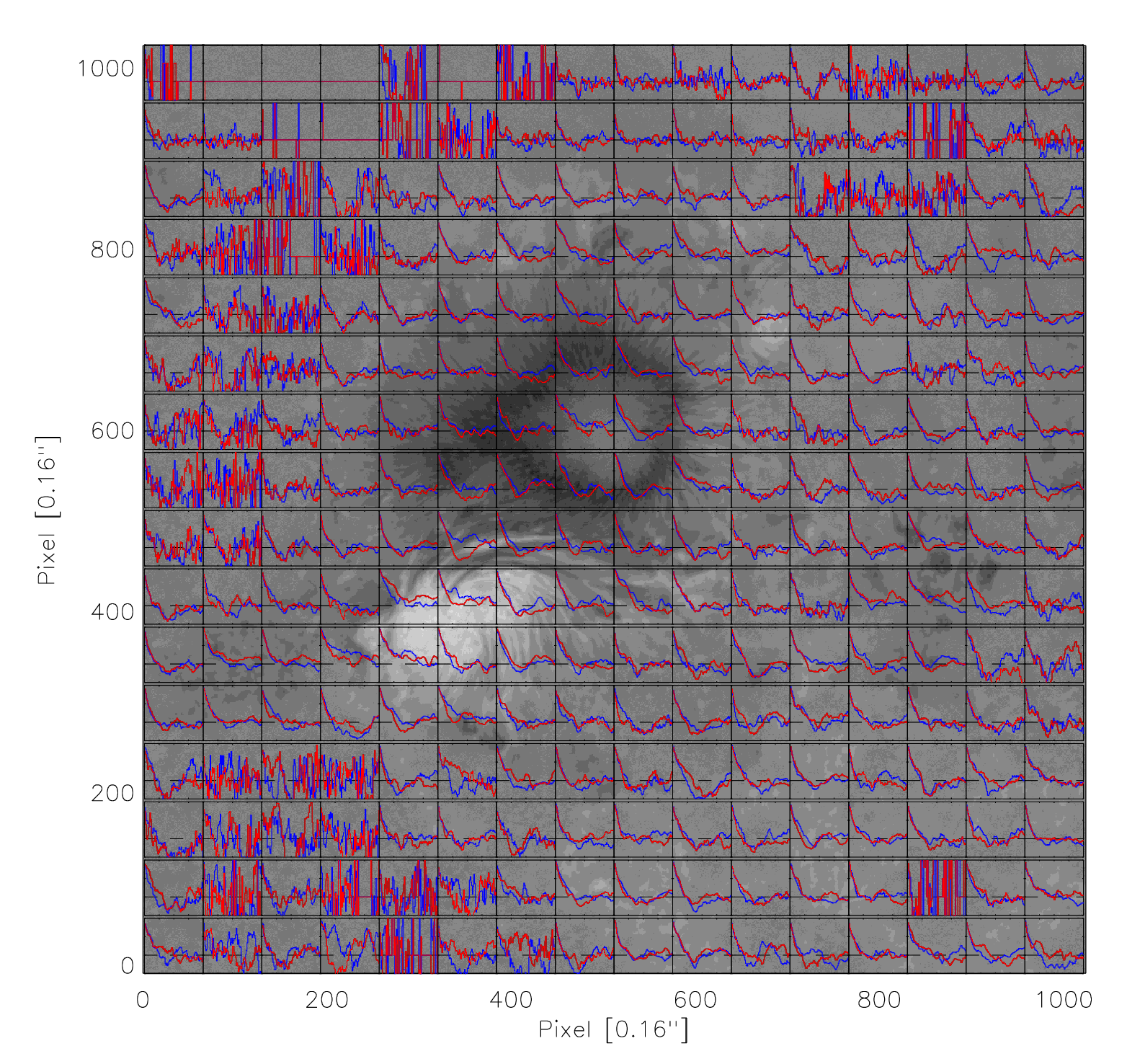,width=6.5in}}
  \caption[]{\footnotesize \textsl{The grayscale image is $|B_{\rm
        LOS}|$ in a full-resolution magnetogram, with saturation at
      $\pm 4$ kG. Autocorrelations of $u_x$ (red) and $u_y$ (blue) in
      subregions of the full field of view.  The flows were derived
      with $\Delta t =$ 64 min and $\sigma = 4$ pix with $(2 \times
      2)$ binned data.  The vertical range in each cell is [-0.5,
        1.0], with a dashed black line at zero correlation.
      Autocorrelations are essentially random in weak-field regions,
      in which few pixels were above the threshold for tracking.}}
        \label{fig:flow_256}
\end{figure}

In Figure \ref{fig:flow_256}, we show autocorrelations of $u_x$ and
$u_y$ in $(32 \times 32)$-pixel subregions of magnetograms with 0.32
\arcsec pixels (so binned $2 \times 2)$.  These flows were derived
with $\Delta t =$ 64 minutes and $\sigma = 4$ pixels. Background gray
contours in the figure show 50 G and 200 G levels of $|B_{\rm LOS}|$
from the initial magnetogram (in full-resolution pixels), so
variations in flow autocorrelations can be understood in the context
of active region structure.

In weak-field regions, few pixels are above the threshold for
tracking, so very few velocities are estimated. Consequently, the
autocorrelations are essentially random.  We therefore only fitted the
autocorrelations in high-occupancy subregions, which we define to be
those with a median (over the dataset's duration) occupancy of pixels
above the tracking threshold of at least 20\%, which gives large
enough samples that the autocorrelations are stable.  The fits to the
magnetic field autocorrelation curves in Figure \ref{fig:mag_256}
convey how well the single-parameter exponential fits match the
autocorrelations for a range of lifetimes and autocorrelation
profiles.

\begin{figure}[ht]
  \centerline{\psfig{figure=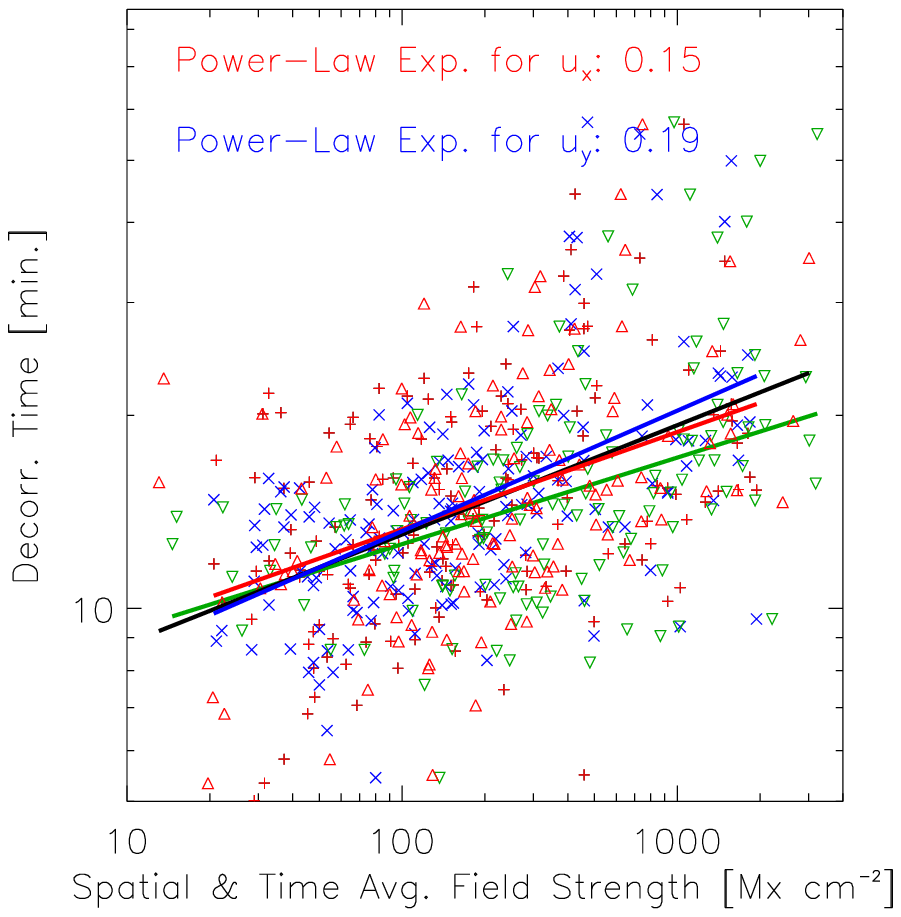,width=5.0in}}
  \caption[]{\footnotesize \textsl{Fitted flow lifetimes for
      non-bubble subregions of Figure \ref{fig:flow_256} in which at
      least 20\% of the $32^2$ $(2 \times 2)$ pixels exceeded the
      tracking threshold, as a function of the spatial average of the
      unsigned, boxcar-averaged LOS magnetic field in each subregion.
      Lifetimes for $u_x$ in subregions are shown with red +'s, $u_y$
      are blue $\times$'s, and fits to these are red and blue solid
      lines.  The red $\Delta$'s (green $\nabla$'s) are lifetimes of
      $u_x$ versus subregion-averaged $|B_z|$ ($|\bvec|$) from the SP
      data, and the black (green) line is a fit, with slope 0.17
      (0.13).  The correlation between flow lifetime and field
      strength is statistically significant, though the fitted
      power-law dependence is weak.}}
        \label{fig:flow_life_vs_b}
\end{figure}

In Figure \ref{fig:flow_life_vs_b}, we show fitted flow lifetimes
($u_x =$ red +'s and $u_y =$ blue $\times$'s) for high-occupancy
subregions of Figure \ref{fig:flow_256}, as a function of the spatial
average of the unsigned, boxcar-averaged LOS magnetic flux density in
each subregion.  Red $\Delta$'s and green $\nabla$'s are lifetimes of
$u_x$ components of the flow as functions of spatial averages of
$|B_z|$ and $|\bvec|$, respectively, and red dashed and green solid
lines are fits to each.  Most flow lifetimes are between about 7 and
30 minutes, confirming the expectation based upon the red, dashed
curves in Figure \ref{fig:bigdt_autocorr} that these flows are
displacement-dominated.  There are statistically significant
correlations both between flow lifetime and field strength, and
between flow lifetime and median occupancy of above-threshold pixels.
The rank-order correlation of field strength $|\bvec|$ with $u_x$ (and
$u_y$) flow lifetime, however, is a statistically significant 0.52
(and 0.51) for the 35 of 137 non-bubble subregions of SP data with
occupancies of 1000 or more (of a possible 1024), implying the
lifetime versus field-strength trend is independent of any
occupancy-lifetime correlation.  Linear fits to logarithms of lifetime as a
function of field strengths return weak power-law exponents, below
$\sim 0.2$ for each flow component.  For flows with $\Delta t =$ 8
minutes and $\sigma = $4 pixels, correlations of lifetime with field
strength are quite weak. For flows estimated with $\Delta t =$ 256
minutes and $\sigma = $16 pixels, fitted power-law exponents between
lifetime and field strengths $\{|B_{\rm LOS}|, |B_z|, |\bvec|\}$ range
from $0.2 -- 0.3$, but field strength-lifetime correlations are weaker
for $u_x$ than for $u_y$.  The east-west orientation of the PIL,
combined with apparent rapid motion of vertical magnetic flux along
the PIL \citep{Demoulin2003}, could produce more rapid evolution of
east-west flow components than north-south flow components.

Correlations between flow lifetime in a subregion and magnetic field
strength in that subregion imply that Lorentz forces drive a component
of photospheric evolution.  To the extent that convection is the
primary driver of photospheric evolution, and magnetic fields inhibit
convection, there is no obvious reason why flow lifetimes should be
longer in stronger-field regions.  But because magnetic structures
persist for longer timescales than flows, it is likely that forces
acting on those fields, such as magnetic buoyancy \citep{Parker1955}
or magnetic torques \citep{Longcope2000} are also longer-lived.
Consequently, it is physically reasonable that in stronger-field
regions, magnetic evolution is partially driven by longer-acting
forces than the pressure gradients arising from convection. 

\begin{figure}[ht]
  \centerline{\psfig{figure=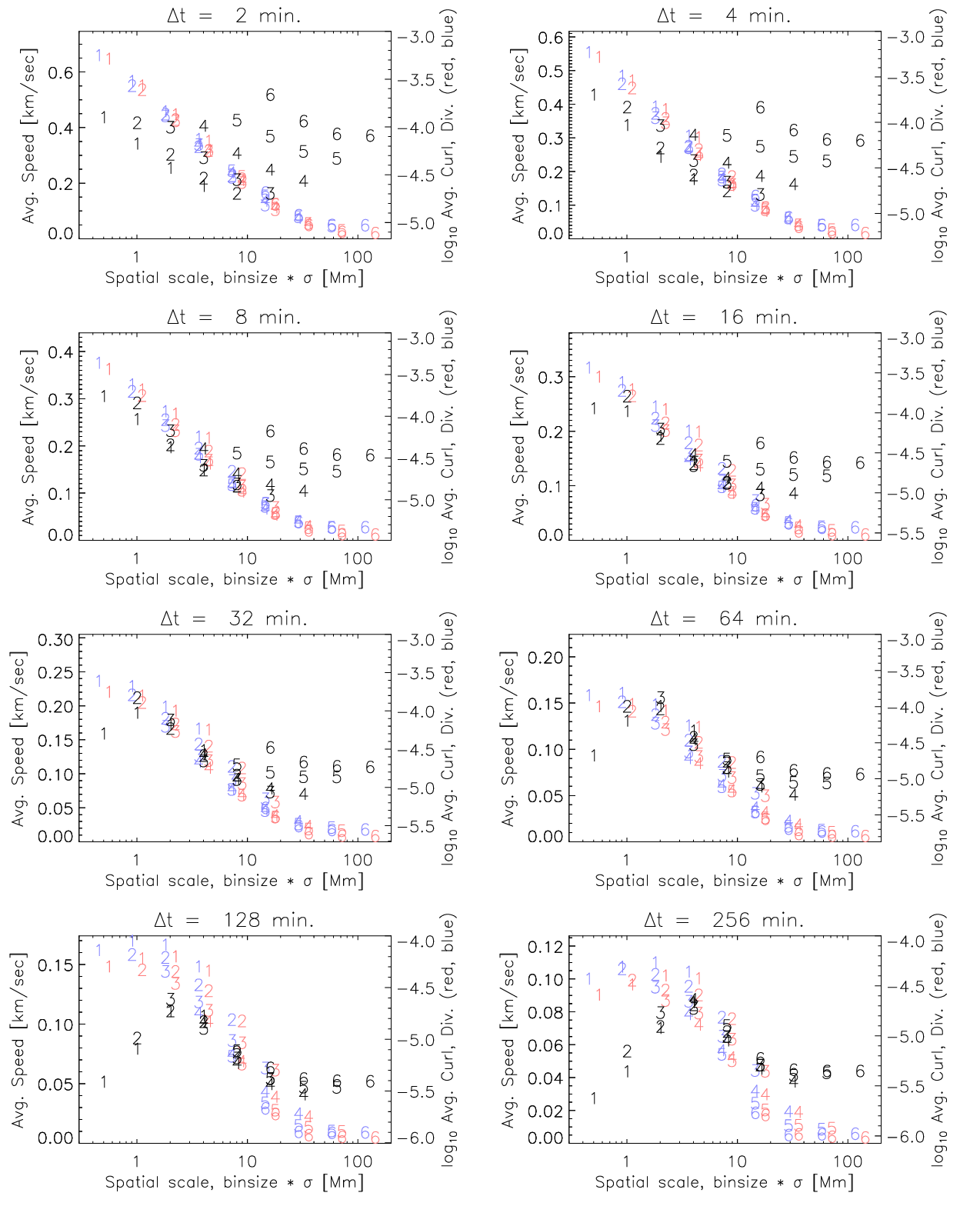,width=6.5in}}
  \caption[]{\footnotesize \textsl{Black numerals show average flow
      speed (left axis) vs. spatial scale, parametrized by the product
      of the macropixel bin size and the windowing parameter $\sigma$,
      for several flow cadences; the numbers correspond to
      $\log_2$(bin size), in units of 0.16 \arcsec pixels.  Red (blue)
      numerals show the log of the average normal curl (horizontal
      divergence) of the flow, on the right scale, and are slightly
      offset right (left) for clarity.  Because we estimated flows for
      several macropixel sizes and $\sigma$'s, we have multiple
      samples at most resolutions.  For shorter $\Delta t$, bin size
      affects the average flow speed at a fixed spatial scale; for
      longer $\Delta t$, bin size does not strongly affect average
      flow speed at a fixed spatial scale.  In contrast, binning seems
      to affect average flow curls and divergences more at larger $\Delta
      t$.}}
        \label{fig:vav_vs_scale}
\end{figure}
%

How does flow lifetime $\tau$ depend upon flow spatial scale?  A crude
estimate of the turnover time of a convective eddy --- expected to be
related to flow lifetime --- is the eddy length scale $L$, divided by
a characteristic speed $u_0$.  As noted by \citet{Berger1998},
however, ``the term `characteristic speed' must be used only in
relation to a given spatiotemporal resolution.''  So as a preliminary
step in characterizing flow lifetime versus spatial scale, we first
investigate flow speed versus spatial scale.  What do our results
reveal about typical flow speeds over varying spatial scales?  We
define a flow's spatial scale as the product of macropixel bin size
and the windowing parameter $\sigma$.  As shown in Figure
\ref{fig:binsigma}, changes in both parameters seem to have similar
effects upon flow lifetimes.  Because we estimated flows for several
macropixel sizes and choices of $\sigma$, we therefore have multiple
sequences of flow estimates at a few spatial resolutions for each
cadence $\Delta t$. We computed the average flow speed at each scale
over all pixels and time steps for which velocities were estimated.
In Figure \ref{fig:vav_vs_scale}, we show average velocity as a
function of spatial scale; the plot symbols are $\log_2$(bin size), in
units of the data's native 0.16 \arcsec pixels, to better illustrate changes
in average velocity as a function of bin size at a given spatial
scale.  (Values of curls and divergences of the flow pattern are also
shown in this figure, and are discussed below.)  One basic property
is evident: average speeds are slower for longer tracking intervals
$\Delta t$, as was noted by \citet{Chae2004b} and \citet{Verma2011}
for both relatively long and short $\Delta t$, respectively. (This is
discussed further below.)  For $\Delta t$ shorter than 16 min, average
speeds peak at the small end of the range of spatial scales, as found
in previous work \citep{Berger1998, Chae2004b, Verma2011}. For longer
$\Delta t$, however, average speeds peak at higher spatial scales:
speeds at small-scales for large $\Delta t$ are low.  The large
$\Delta t$ implies that small-scale structures at $t_i$ and $t_f$ have
evolved substantially.  While it is plausible that FLCT would infer
random displacements in this case, and that averaging over these
random displacements (all positive) would produce a large mean speed,
we do not observe this.  Apparently, for a given $\Delta t$, the FLCT
code is primarily sensitive to flows on a particular spatial scale.
Also, for shorter $\Delta t$, average speed at a given spatial
resolution depends strongly upon spatial binning, while for longer
$\Delta t$, average speeds seem insensitive to spatial binning.  This
probably reflects different effects from the combination of actual
evolution and noise on the apparent magnetic evolution convolved with
spatial binning and apodization by $\sigma$.  For instance, large
values of $\sigma$ presumably average over the effects of actual
evolution on sub-$\sigma$ scales.  If so, speeds should be higher with
smaller $\sigma$ at a given spatial resolution, as we find.
We remark that the combined effects of cadence with binning and
apodization preclude straightforward testing our FLCT velocities for
consistency with the Kolmogorov scaling (e.g.,
\citealt{Thompson2006b}, p.68).
%

We note that tests of FLCT with synthetic magnetograms from MHD
simulations, in which the true velocities were known, showed that FLCT
systematically underestimates speeds, by $\sim 25$\%
\citep{Welsch2007}, so the absolute values of the average speeds we
find here are probably systematically lower than estimates that would
be found using other tracking codes.  Further, because our velocities
are derived from magnetic evolution, and magnetic flux both tends to
inhibit convection and to concentrate in stable downflow regions
between convective cells, it is reasonable to expect our average
speeds to be systematically lower than tracking results derived
from intensity images.  We note that for $\Delta t$ of 2 or 4 min, our
average speeds at the smallest spatial scale are close to those
reported by \citet{Verma2011} (see their Table 3).  The speeds of FLCT
flows derived by \citet{Welsch2009}, with $\Delta t$ of 96 min and a
spatial scale of $\sim$12 Mm were predominantly in the 50 -- 100 m
sec$^{-1}$ range (see their Fig. 6), comparable to values we find for
similar $\Delta t$ and spatial scales here.

\begin{figure}[ht]
  \centerline{\psfig{figure=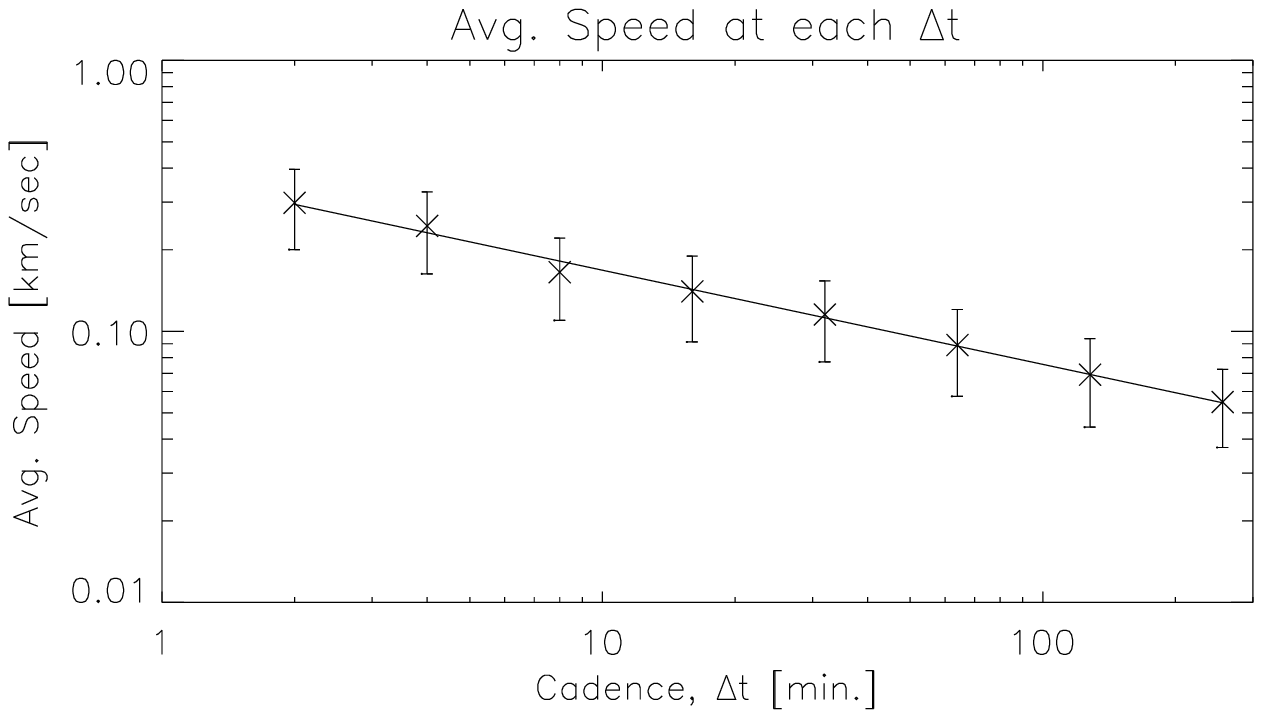,width=5.5in}}
  \caption[]{\footnotesize \textsl{Flow speed as a function of
      cadence, averaged over both space and time for each spatial
      resolution (binning $\times \sigma$) and over spatial
      resolutions at each $\Delta t$.  Error bars show standard
      deviation over resolutions.  Least squares linear fit to the
      logarithms of cadence and speeds returns a power-law exponent
      of -0.34.}}
        \label{fig:vav_vs_dt}
\end{figure}

In Figure \ref{fig:vav_vs_dt}, we show the average speed versus
tracking cadence $\Delta t$, averaged over both space and time in all
flow maps at a given spatial resolution and all spatial resolutions at
each cadence.  Effectively, this amounts to computing the average of
speeds at each $\Delta t$ in Figure \ref{fig:vav_vs_scale} --- fitting
the speeds in each panel to a horizontal line.  Error bars in Figure
\ref{fig:vav_vs_dt} reflect the standard deviation in speeds in each
panel of Figure \ref{fig:vav_vs_scale}.  A linear fit to the
logarithms of cadences and average speeds returns an power-law
exponent of -0.34.  \citet{Verma2011} computed average speeds for a
range of much shorter $\Delta t$'s, and over the eight-fold increase
in $\Delta t$ from 60 to 480 sec, their average speed fell by about a
factor of two. This implies a scaling near -1/3, similar to ours.
(But they found an {\em increase} in average speed between $\Delta
t$'s of 15 and 60 sec, and similar average speeds at 60 and 90 sec.)
They suggest decreasing speeds with increasing $\Delta t$ might arise
from the short lifetimes of features in the images they
track. However, we find the same effect when tracking much longer-lived
magnetogram structures, implying the limited lifetimes of features is
probably not solely responsible for decreasing speeds with increasing
$\Delta t$.  
We also note that both \citet{Hagenaar2005} and \citet{Verma2011b} found
that individual, successive moving magnetic features (MMFs) followed
preferential paths as they moved outward from sunspots, suggesting
short-lived features can reveal longer-lived flow patterns.
As noted above, a longer $\Delta t$ can average over the
effects of short-lived flows: displacements inferred with a longer
$\Delta t$ might not recover zig-zag motions that can be resolved with
shorter $\Delta t$, with the shorter apparent path length at longer
$\Delta t$ implying a lower estimate of speed.  

\citet{Verma2011} also averaged higher-cadence flow maps over
increasing time periods --- 1, 2, 4, 8, and 16 hr --- and found
statistics of averaged-flow speeds (mean, median, 10th percentile,
maximum) all systematically decreased.  Such averaging is not the same
as tracking over longer $\Delta t$, as we have done, but the trends
are qualitatively similar: a log-log plot of the averaging periods and
mean speed from tracking granulation given in Table 1 of
\citet{Verma2011} also shows a power-law decrease of mean speed with
time between 1 -- 8 hr, though a linear fit to the logarithms returns
an exponent near -0.18, about half our result.  It is plausible that
short-lived flows, resolved with shorter $\Delta t$, would primarily
reflect the essentially random effects of granulation, and therefore
average incoherently over long times.  However, comparison of our
large-$\Delta t$ flow speeds with speeds inferred by \citet{Verma2011}
with a shorter tracking interval $\Delta t'$ and then averaged over
longer time intervals similar to ours implies that shorter-timescale
flows are averaging coherently.  It is possible that the motion
tracers --- either the intensity pattern, or magnetic flux --- used in
one approach or the other (or even both!) do not faithfully represent
the true plasma velocity. In this vein, \citet{Rieutord2001} has
argued that the motion of granules does not accurately reflect
underlying velocities on spatial and temporal scales that are too
short.
This suggests flows inferred from magnetic tracking by flow maps
derived with short tracking intervals but averaged over long $\Delta
t$ and flow maps derived with a tracking interval of $\Delta t$ should
be compared in a future study.  
%

%
\begin{figure}[ht]
  \centerline{\psfig{figure=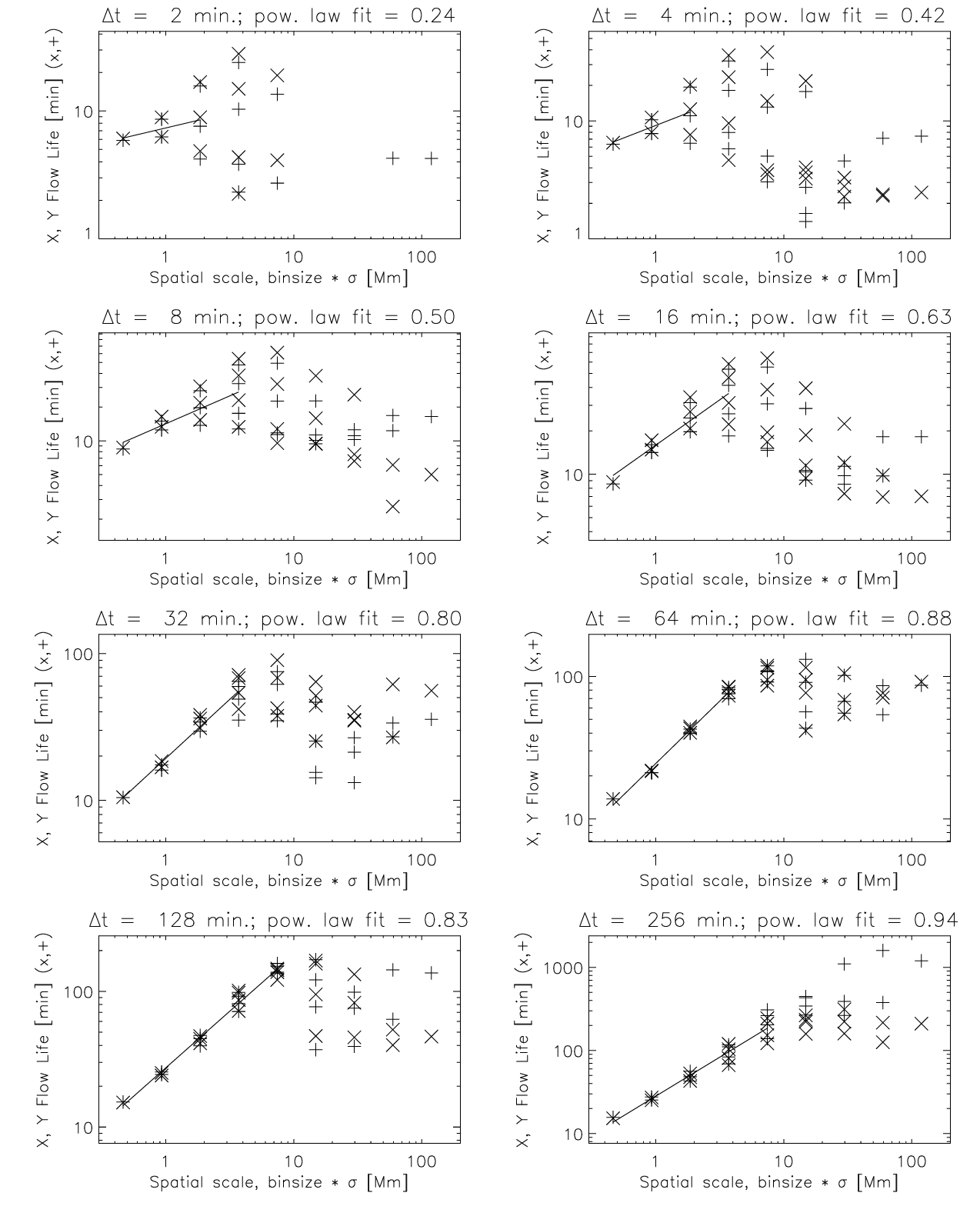,width=6.5in}}
  \caption[]{\footnotesize \textsl{Flow lifetimes (+'s for $u_x$,
      $\times$'s for $u_y$) as a function of flow spatial scale, as
      parametrized by the product of the macropixel bin size and the
      windowing parameter $\sigma$, for several flow cadences.
      Because we estimated flows for several macropixel sizes and
      $\sigma$'s, we have multiple samples at some resolutions.  We
      computed linear fits (solid lines) to the logarithms of spatial
      scale and lifetime over a limited range of scales.  The fitted
      power-law exponent is printed in the title of each cadence's
      plot.}}
        \label{fig:flow_life_vs_scale}
\end{figure}
%

We now return to the question of how flow lifetime $\tau$ varies with
spatial scale.  As above, we define the spatial scale of an estimated
flow as the product of macropixel size and windowing parameter
$\sigma$, and therefore have multiple sequences of flow estimates at a
few spatial resolutions for each cadence $\Delta t$.  For each
sequence of flow estimates, we can estimate the flow lifetime $\tau$
over the whole active region from a one-parameter fit to the
autocorrelation as function of lag time $\Delta T$, assuming
exponential decay, $\exp{(-\Delta T/\tau)}$. In Figure
\ref{fig:flow_life_vs_scale}, we plot flow lifetimes estimated this
way, as a function of spatial scale, for several tracking cadences
$\Delta t$.  In the shorter-scale region of each log-log plot, flow
lifetime seems to scale, roughly, in a straight line with spatial
scale.  To characterize this dependence, we fit power-law exponents
for each cadence over the range of seemingly straight-line log-log
scaling (shown with a solid line on each plot).  For longer cadences,
flow lifetime appears to scale nearly linearly with scale length,
while the dependence of flow lifetime on length scale seems weaker for
shorter cadences.  At larger length scales in each plot, the
relationship between lifetime and length scale becomes less coherent
as binning and apodization are varied.  \citet{Welsch2009} found
lifetimes at $\sim$ 12 Mm to be two to three steps at their $\Delta t$
= 96 min magnetogram cadence.  For similar spatial scale and $\Delta
t$ (64 and 128 min) we find shorter lifetimes, by a factor of two or
so, although this is also the spatial scale where our lifetime
estimates become incoherent.
For all spatial scales we investigate, the flow lifetime is shorter
than that predicted from a linear scaling of lifetime with length
scale, sometimes dramatically so.
We note that our estimates of lifetimes are rough
approximations. Longer-lived flows (e.g., supergranular flows) would
remain undetected in our lifetime estimates if their asymptotic
autocorrelation lies below our $1/e$ cutoff.
The longer lifetimes of larger-scale flows is another possible
explanation of the correlation between flow lifetime and magnetic
field strengths: stronger magnetic fields might couple nearby regions
of plasma more strongly (making the plasma more rigid), leading to
larger spatial scales for flows in more strongly magnetized regions;
and such flows should persist for longer than smaller-scale
flows.

\begin{figure}[ht]
  \centerline{\psfig{figure=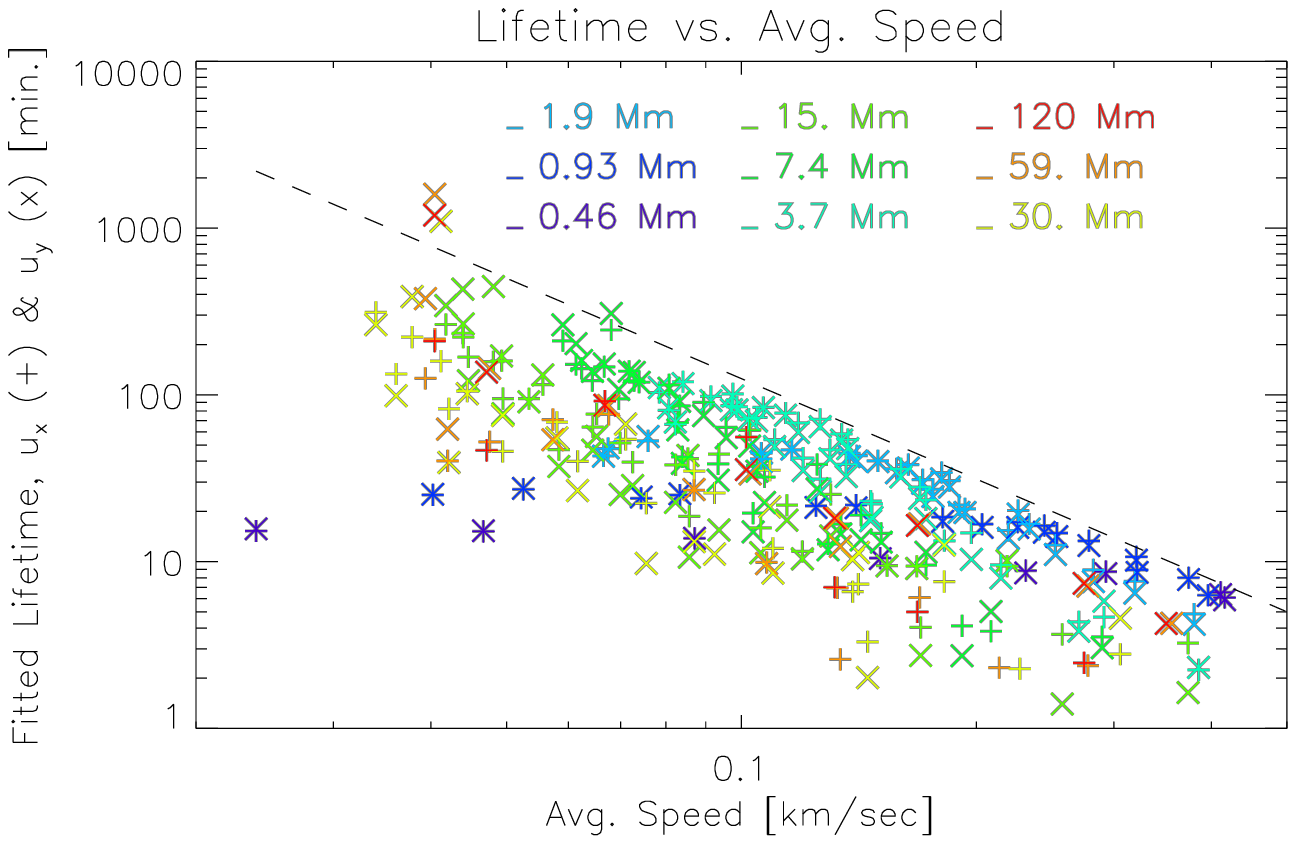,width=5.5in}}
  \caption[]{\footnotesize \textsl{Fitted lifetime vs. flow speed
      averaged over space and time for each choice of $\Delta t$,
      spatial binning, and $\sigma$.  Lifetimes for $u_x$ and $u_y$
      are plotted with $+$'s and $\times$'s, respectively. Generally,
      higher average speeds correspond to shorter lifetimes.  While a
      range of lifetimes exists at each average speed, there appears
      to be an upper limit at a given average speed, with the peak
      fitted lifetime scaling roughly as (average speed)$^{-2}$
      (dashed line; note that this line is not a fit).  Points are
      color-coded by spatial scale of the flow (binning
      $\times \sigma$). }}
        \label{fig:life_vs_vav}
\end{figure}

In Figure \ref{fig:life_vs_vav}, we show fitted flow lifetimes for $x$
and $y$ components of the flow (denoted $+$ and $\times$,
respectively) as functions of $\langle v \rangle$, the flow speed
averaged over space and time for each combination of tracking interval
$\Delta t$, spatial binning, and apodization parameter $\sigma$.  Plot
symbols are color-coded by spatial scale, defined as the product of
spatial binning and apodization parameter.  Generally, higher average
speeds correspond to shorter lifetimes.  While a range of lifetimes
was found at each average speed, there appears to be a rough upper
limit at a given speed, with the peak fitted lifetime $\tau_*$ scaling
roughly as $\langle v \rangle^{-2}$.  Noting that the product of a
timescale and the square of speed yields a quantity with the
dimensions of kinematic viscosity (length$^2$/time), the mean and
standard deviation of $\tau_* \langle v \rangle^2$ are $(2.7 \pm 1.9)
\times 10^{11}$ cm$^2$ sec$^{-1}$ and $(2.9 \pm 2.3) \times 10^{11}$
cm$^2$ sec$^{-1}$ for the $x$ and $y$ components of the flow,
respectively.  The limit of flow lifetimes shown by the dashed line in
Figure \ref{fig:life_vs_vav}, $1.25 \langle v \rangle^{-2}$, would
correspond to a kinematic viscosity of $7.5 \times 10^{11}$ cm$^2$
sec$^{-1}$.  While the physical significance of this value is unclear
(if it is even physical --- it might be an artifact of our methods),
it roughly agrees with values of viscosity near the photosphere
reported by \citet{Nesis1990}.

%


\subsection{Lifetimes of Curls \& Divergences}
\label{subsec:curldiv}


%
\begin{figure}[ht]
  \centerline{\psfig{figure=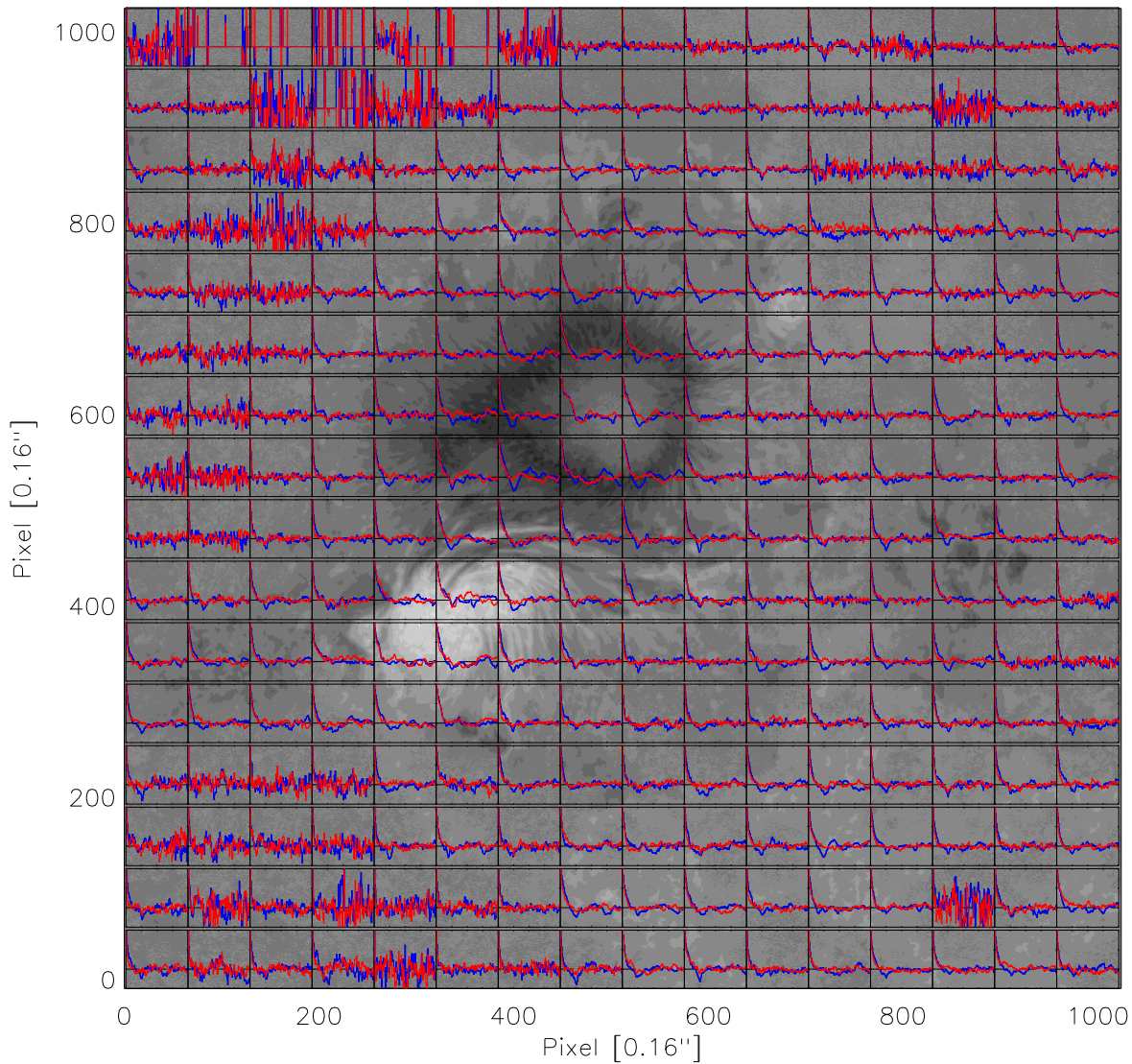,width=6.5in}}
  \caption[]{\footnotesize \textsl{The grayscale image is $|B_{\rm
        LOS}|$ in a full-resolution magnetogram, with saturation at
      $\pm 4$ kG. Autocorrelations of normal curls (red) and
      horizontal divergences (blue) in each subregion of the full
      field of view are overplotted. The flows were derived with
      $\Delta t =$ 64 min and $\sigma = 4$ pix. The vertical range in
      each cell is [-0.5, 1.0], with a dashed black line at zero
      correlation. Autocorrelations are essentially random in
      weak-field regions, in which few pixels were above the threshold
      for tracking.}}
        \label{fig:curldiv_varbin}
\end{figure}
\begin{figure}[ht]
  \centerline{%
\psfig{figure=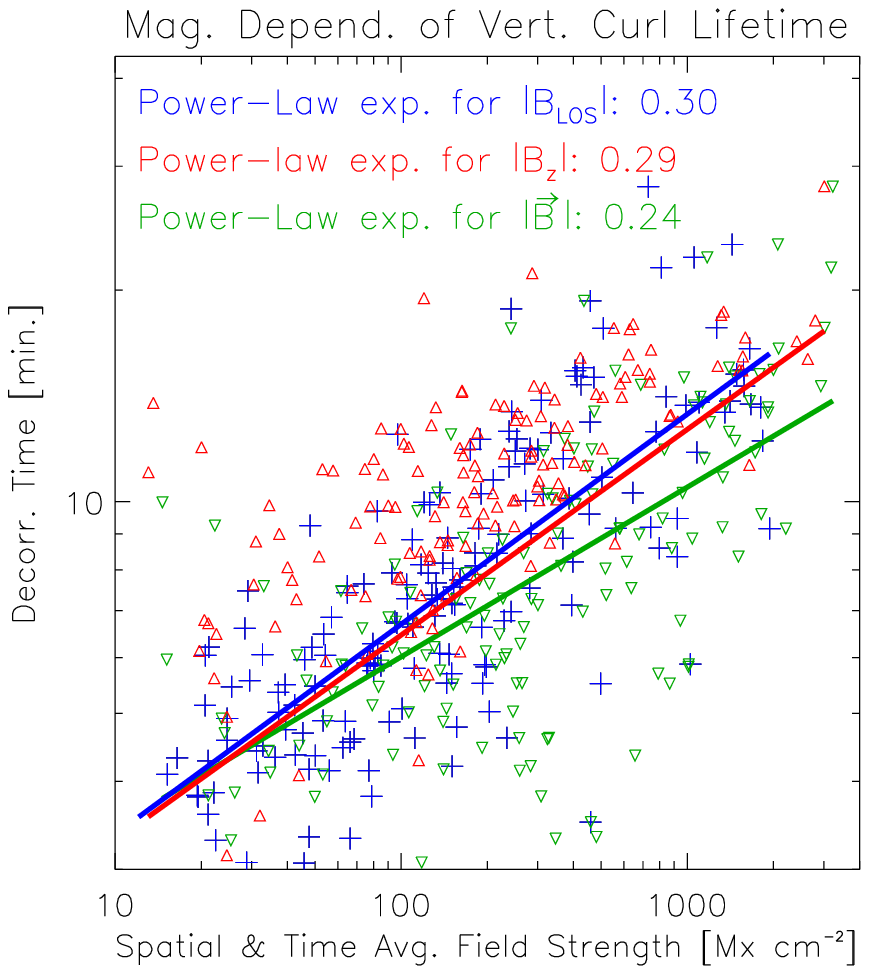,width=3.25in}%
\psfig{figure=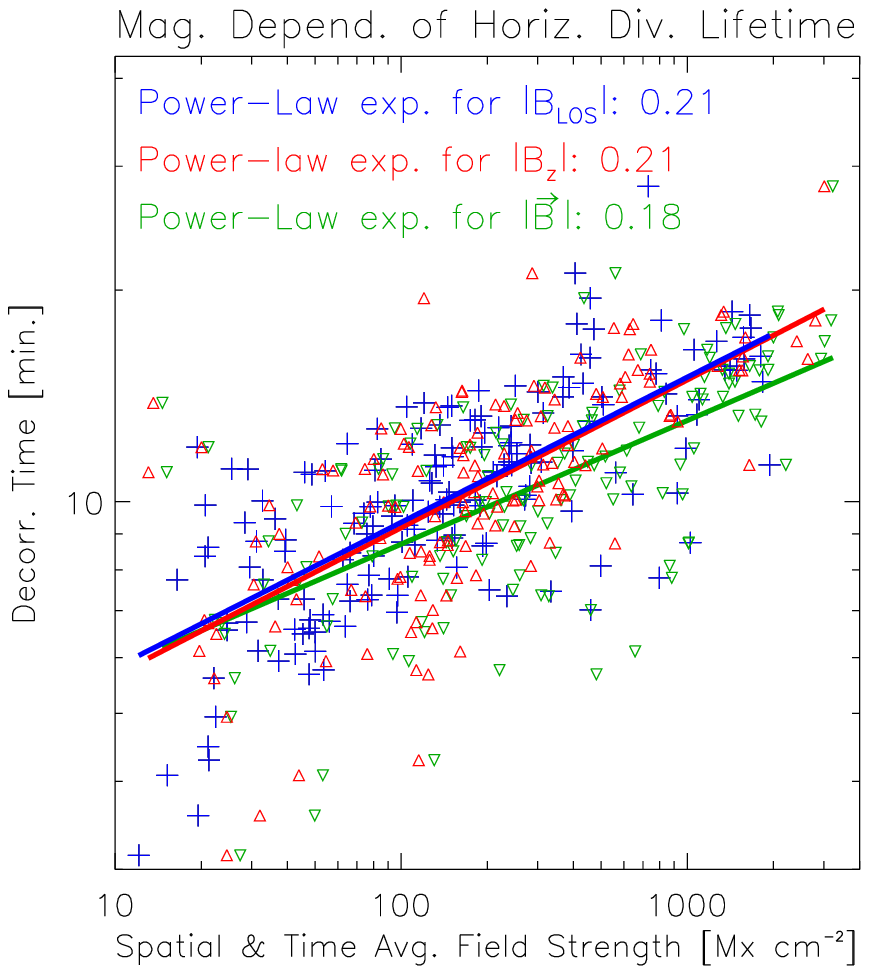,width=3.25in}}
  \caption[]{\footnotesize \textsl{Fitted lifetimes of normal curls
      (left) and horizontal divergences (right) for subregions of
      Figure \ref{fig:curldiv_varbin} in which at least 20\% of the
      $32^2$ $(2 \times 2)$ pixels exceeded the tracking threshold, as
      functions of the spatial average $|B_{\rm LOS}|$ (from NFI; blue
      +'s), $|B_z|$ (from SP; red $\Delta$'s), or $|\bvec|$ (from SP;
      green $\nabla$'s) in each subregion.  The correlation between
      curl/divergence lifetime and field strength is statistically
      significant, though the fitted power-law dependence is weak.
      The magnetic dependence of the lifetimes of curls is somewhat
      stronger than that of lifetimes of divergences.}}
        \label{fig:curldiv_vs_b}
\end{figure}

The lifetimes of normal curls and horizontal divergences of the flow
fields, defined as
\bea \hatn \cdot (\nabla_h \times \uvec) \\ 
(\nabla_h \cdot \uvec) ~, \label{eqn:curldiv} \eea
respectively, are also of interest: vortical flows can efficiently
inject magnetic energy and helicity (e.g., \citealt{Longcope2000}),
and diverging flows are associated with flux emergence in both
simulations \citep{Abbett2000} and observations \citep{Welsch2009}.
In Figure \ref{fig:vav_vs_scale}, we have overplotted mean absolute
values of curls (divergences), averaged over tracked pixels in $x-y-t$
datacubes of our flow maps, in red (blue), computed using finite
difference approximations with the smallest binning, twice the
native 0.16 \arcsec pixel size, in the denominator.  For consistency with
our definition of spatial scale used above, the denominator should be
the product of macropixel size and apodization parameter $\sigma$.
This would, however, introduce an additional inverse scaling with the
variable plotted on the horizontal axis.  Average values of curls and
divergences at our smallest length scales for short $\Delta t$ are on
the order of a few times 10$^{-4}$ sec$^{-1}$, similar to values
reported by \cite{Verma2011}.  For fixed $\Delta t$, curls and
divergences fall by roughly two decades over the approximately two
decades in length scales we study, implying curl and divergence
magnitudes scale approximately as 1/length.  Average magnitudes of
curls and divergences decreased by about a decade over the range of
$\Delta t$'s in our study, a faster decrease with $\Delta t$ than for
speeds.

In Figure \ref{fig:curldiv_varbin}, we show autocorrelations of normal
curls (red) and horizontal divergences (blue) computed in subregions
of the full FOV, superimposed on 50 G and 200 G contours of $|B_{\rm
  LOS}|$, for flows estimated from 2$^2$-binned data, with $\sigma =
4$ macropixels and $\Delta t$ = 64 min.  In Figure
\ref{fig:curldiv_vs_b}, we show the fitted lifetimes for curls (left)
and divergences (right) versus subregion- and boxcar-averaged $|B_{\rm
  LOS}|$ from NFI, and subregion-averaged $|B_z|$ and $|\bvec|$ from
SP, where, as above, lifetimes were determined by one-parameter fits
to subregions' autocorrelation functions assuming exponential decay.
As with the lifetimes of $x$ and $y$ components of the flow, curl and
divergence lifetimes exhibit a dependence on both field strength and
the median occupancy of above-threshold pixels in each 32$^2$
subregion of $2 \times 2$-binned full-resolution pixels.
The rank-order correlation of field strength $|\bvec|$ with curl (divergence)
lifetime, however, is a statistically significant 0.59 (0.58) for the
59 subregions with occupancies of 1000 or more (of a possible 1024),
implying the lifetime versus field-strength trend is independent of 
any occupancy-lifetime correlation.
We repeat that our estimates of $|B_{\rm LOS}|$ suffer from
non-monotonicity in the Stokes $V/I$ ratio, while our spatial averages
of $|B_z|$ and $|\bvec|$ ignore the likelihood of small filling
factors for magnetic field structures.
Despite these limitations, the existence of a trend seems clear, and
we find a stronger dependence of the lifetimes of curls on field
strength than for lifetimes of divergences.
We note that, as with the lifetime of flows themselves, the fitted
dependence of lifetime of curls and divergences on $|\bvec|$ appears
weaker than that on either $|B_z|$ or $|B_{\rm LOS}|$.

\begin{figure}[ht]
  \centerline{\psfig{figure=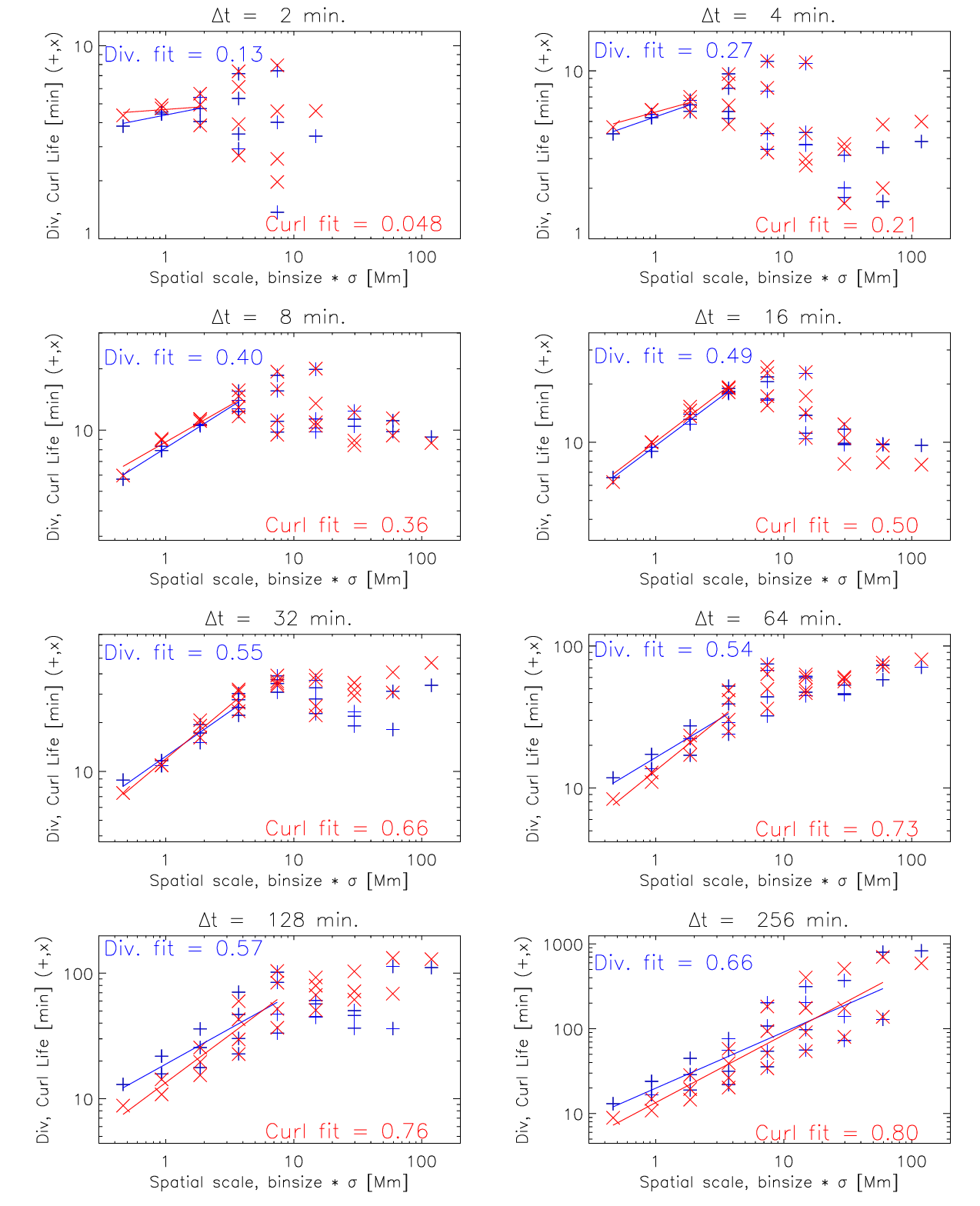,width=6.5in}}
  \caption[]{\footnotesize \textsl{Lifetimes of normal curls (red
      $\times$'s) and horizontal divergences (blue +'s) as a function
      of flow spatial scale, as parametrized by the product of the
      macropixel bin size and the windowing parameter $\sigma$, for
      several flow cadences.  Because we estimated flows for several
      macropixel sizes and $\sigma$'s, we have multiple samples at
      some resolutions.  We computed linear fits (solid lines) to the
      logarithms of spatial scale and lifetime over a limited range of
      scales, and list the fitted power-law exponents on each
      cadence's plot.}}
        \label{fig:curl_life_vs_scale}
\end{figure}
Finally, in Figure \ref{fig:curl_life_vs_scale}, we show the lifetimes
of normal curls (red $\times$'s) and horizontal divergences (blue +'s)
versus spatial scale, defined as above as the product of macropixel
bin size and windowing parameter $\sigma$, for each tracking interval
$\Delta t$.  Because we estimated flows for several macropixel sizes
and $\sigma$'s, we have multiple samples at some resolutions, most of
which roughly agree.  At most cadences, the lifetimes increase with
spatial scale, and then saturate.  In Figure
\ref{fig:curl_life_vs_scale}, we also show linear fits to the
logarithms of lifetimes as functions of logarithms of spatial scales
over the range of spatial scales before saturation occurs.

\section{Summary \& Discussion}
\label{sec:discussion}

We used FLCT to estimate flows from a $\simeq 13$ hour sequence of
high-cadence ($\simeq 2$ min), high-resolution (0.16 \arcsec pixel,
0.3 \arcsec diffraction limit) line-of-sight magnetograms of NOAA AR
10930.  This seeing-free, high-resolution, long-duration dataset
enabled us to conduct a detailed study of the effects of tracking
parameter selection on the estimated flows, and to investigate the
lifetimes of flows and flow structure (normal curls and horizontal
divergences) as functions of spatial scale and magnetic field
strength.  For context, we also characterized the lifetimes of
magnetic structures.

Several conclusions can be drawn from our results, which may be
divided into practical conclusions relevant to the process of flow
estimation from magnetograms, versus physical conclusions about
surface magnetic flows and evolution.  Our practical conclusions are:

\blb

\item Flows estimated by magnetic tracking (and related optical flow)
  methods with a particular set of tracking parameters (i.e., tracking
  interval $\Delta t$ between images, pixel size, apodization window
  size $\sigma$) are primarily sensitive to evolution on a particular
  spatial scale and time scale.

\item If the tracking interval $\Delta t$ between pairs of
  magnetograms is too short, then changes in the field between
  magnetograms is mostly due to noise instead of actual magnetic
  evolution.  This is the noise-dominated regime.  Its hallmark is
  rapid fluctuations in estimated velocities, which can be quantified
  by autocorrelating maps of the velocity components.  The effect of
  noise can be reduced by: increasing $\Delta t$, averaging successive
  magnetograms prior to tracking, or both.

\item Because magnetic features persist for much longer than the
  intensity features to which tracking methods have also been applied
  (e.g., \citealt{Berger1998, Verma2011}), the upper limit for $\Delta
  t$ is not strongly constrained in magnetic tracking.  If, however,
  the tracking interval $\Delta t$ between pairs of magnetograms is
  longer than the flow lifetime at that spatial scale, then the
  estimated flows reflect an {\em average} velocity over more than one
  flow lifetime.  This is the displacement-dominated regime.  Although
  this time-averaged velocity is physically meaningful, it does not
  represent the actual plasma velocity field at a particular time
  (e.g., the center of the tracking interval).

\item Hence, choice of $\Delta t$ can determine whether flow estimates
  are noise-dominated ($\Delta t$ too short), displacement-dominated
  ($\Delta t$ longer than the flow lifetime), or representative of
  actual flows.  To estimate actual plasma velocities, appropriate
  choice of $\Delta t$ can be established through {\em post-facto}
  analysis of flow maps, by autocorrelating flows to find a $\Delta t$
  that both maximizes frame-to-frame autocorrelations (which thus
  minimizes the effects of noise) and is less than the flow lifetime
  $\tau$, which we take to be the time at which the autocorrelation
  drops to $1/e$.

\item Average estimated speeds decrease with increasing tracking
  interval $\Delta t$ between magnetograms (see Figure \ref{fig:vav_vs_dt}).

\el

Our physical conclusions are:

\blb

\item As noted by previous researchers (e.g., \citealt{Berger1998}),
  photospheric flows operate over a range of spatial scales, and exist
  for a range of lifetimes.  Hence, referring to ``the flow field,''
  by itself, is imprecise: explicit reference to spatial and temporal
  scales must also be made.

\item Magnetic structure persists substantially longer in
  stronger-field regions generally, and longer in regions of strong
  vertical field $|B_z|$ than in regions of strong magnetic field
  strength $|\bvec|$.

\item Magnetic structure persists substantially longer when averaged
  over larger spatial scales.  Flow components and their derivatives
  estimated on larger spatial scales also persist for significantly
  longer, though lifetimes scale less than linearly with length scale.

\item Flow components, horizontal divergences, and normal curls
  persist somewhat longer in regions with stronger fields: our
  estimates of their lifetimes are significantly statistically
  correlated with spatially averaged values of each of $|B_{\rm
    LOS}|$, $|B_z|$, and $|\bvec|$.  We find the lifetimes of curls
  exhibit a slightly stronger dependence on field strength than either
  flows components or horizontal divergences.

\item Flows with faster average speeds typically exhibit shorter
  lifetimes.  Also, there appears to be a rough upper limit on flow
  lifetime at a given average speed, which scales as the inverse of
  the square of the average speed.

\el

In this study, we found sets of differing velocity fields for a single
magnetogram dataset by applying a single correlation tracking method
over various spatial and temporal scales.  This is not surprising:
coherent flow structures on varying scales are known to be present in
the photospheric flow field, e.g., global-scale Sun's differential
rotation (in our synodic frame) and meridional flow, high-speed
granular flows and lower-speed supergranular flows driven by
convection, and possibly flows arising from Lorentz forces in
magnetized regions.  In addition, turbulent flows are likely present.
This, however, raises the question, ``Given that a family of velocity
fields can be derived on several scales, how can one determine the
best estimate of the velocity field?''  But what determines ``best?''
It is tempting to try to estimate the ``true'' velocity field in each
resolution element (pixel), but what does this mean physically?
Assuming the photospheric plasma behaves exactly as a fluid down to
some kinetic scale, then within our smallest resolution element there
is a distribution of subresolution velocities, and a meaningful
definition of the ``true'' velocity must be defined in terms of this
distribution.  If the distribution of velocities is peaked in speed
and direction, then some statistical property of each (e.g., mean or
median) might meaningfully quantify the flow.  It is, however,
probable that sub-resolution flows are consistent with a power-law
distribution of flow energy with spatial wavenumber, similar to that
proposed by Kolmogorov (though perhaps with a different wavenumber
scaling), through some inertial range down to a dissipation scale.  A
reasonable expectation based upon our results, as well as those of
previous studies \citep{Berger1998, Verma2011}, is that flows on
successively smaller scales will have higher speeds.  In this case,
what we infer from the motion of magnetic flux (or of brightness
variations, if tracking intensity) is an ``effective flow'' --- the
combined effect of flows acting on all scales.  But it is probable
that the effective flow does not correspond to ``the'' physical
velocity.

From the practical standpoint of understanding how flows are related
to other phenomena of interest (e.g., flares, CMEs, coronal heating,
the dispersal of active region magnetic flux), we must investigate
relationships between these phenomena and flows on varying spatial and
temporal scales to understand which (if any) scales are most relevant.  
%
%
A key challenge is that the largest velocities are present at the
smallest scales: while in a sense these strongest flows are the
easiest to detect, they also have the shortest lifetimes, and are
therefore probably the least relevant for some processes with longer
characteristic time scales, e.g., the emergence of large-length-scale
magnetic flux systems in active regions.  Computing the Poynting flux,
for instance, based upon an instantaneous estimate of flows on
granular scales will likely give an incomplete (if not misleading)
picture of how much and where magnetic energy is being transported
across the photospheric layer.  Hence, much work remains to understand
photospheric flows and their relationships to processes like the
transport of magnetic energy across the photosphere.

We remark that the techniques we have applied here can also be applied
to both real and simulated vector magnetogram sequences, to better
understand evolution of both magnetic fields and flows.  Long
sequences of vector magnetograms (from, e.g., HMI) could be tracked to
analyze flow properties over a range of spatial and temporal scales as
functions of the vector magnetic field's properties.  The lifetimes of
magnetic and flow fields from simulations of photospheric evolution
(e.g., \citealt{Cheung2010}) can also by analyzed via autocorrelation
for comparison with lifetimes inferred from observations.

We expect observations using the newest generation of ground-based
solar telescopes --- NST \citet{Goode2010}, GREGOR
\citep{Volkmer2010}, ATST \citep{Rimmele2010}, and EST
\citep{Zuccarello2011} --- to play a key role in future studies of
flows at high resolution.

\acknowledgements BTW gratefully acknowledges support from the NSF's
SHINE program under award \# ATM-0752597, the NSF's National Space
Weather Program under award \# AGS-1024862, and the Japan Society for
the Promotion of Science.  BTW thanks Ed DeLuca for suggesting parts
of the project.  {\it Hinode} is a Japanese mission developed and
launched by ISAS/JAXA, collaborating with NAOJ as a domestic partner,
NASA and STFC (UK) as international partners. Scientific operation of
the {\it Hinode} mission is conducted by the {\it Hinode} science team
organized at ISAS/JAXA. This team mainly consists of scientists from
institutes in the partner countries. Support for the post-launch
operation is provided by JAXA and NAOJ (Japan), STFC (U.K.), NASA
(U.S.A.), ESA, and NSC (Norway).

\appendix

\section{Calibration by Co-alignment with SP Vector Magnetograms}
\label{app:sp}

\cite{Schrijver2008} used SP data to produce vector magnetograms of AR
10930 over 12 -- 13 Dec. 2006, which are available online.  One of the
magnetograms they produced overlaps with our tracking run.  Here, we
describe the procedure we used to interpolate magnetic field data from
that magnetogram onto a downsampled, cropped grid corresponding to our
NFI observations.  Note that our ultimate goal is to be able to relate
the lifetimes of magnetic field structures and flows over {\em
  regions} of the NFI FOV to average magnetic properties in those
regions.  Hence, we do not seek pixel-scale agreement between the SP
and NFI magnetograms. Rather, we seek approximate agreement on the
scales of a few arcseconds.

\begin{figure}[ht]
  \centerline{\psfig{figure=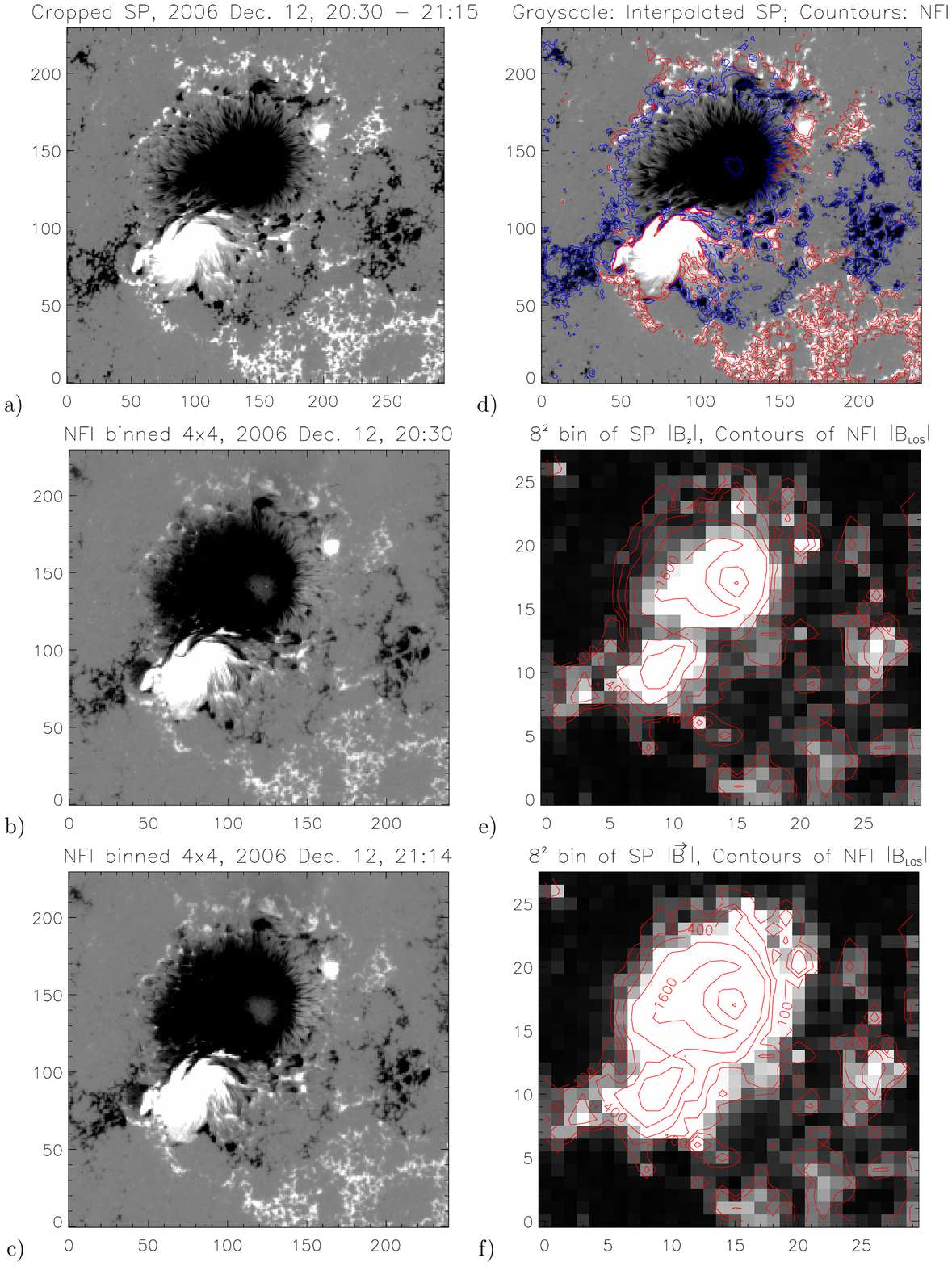,width=5.5in,clip=true}}
  \caption[]{\footnotesize \textsl{Left column: a) SP map of $B_z$,
      scanned over 45 minutes. b) \& c) NFI images binned to $(4
      \times 4)$ from 0.16 \arcsec pixels, from the beginning and end of the
      SP scan. Note lateral extent of SP data is 295 pixels, while NFI
      data are 240 pixels.  Hence, the primary difference between the
      datasets is lateral stretching. Right column: d) Contours of
      $B_{\rm LOS}$ from scan-centered NFI magnetogram ($-/+$ 100G and
      400 G in red/blue, resp.) overlain on the interpolated SP
      data. Slight inaccuracies in co-alignment vary across the
      FOV. e) \& f) $|B_{\rm LOS}|$ from NFI contoured over grayscale
      of $|B_z|$ and $|\bvec|$ from SP, resp., both binned $(8 \times
      8)$. Contours of $|B_{\rm LOS}|$ from NFI correspond to 100,
      200, 400, 800, and 1600 G.}}
        \label{fig:calib_sp}
\end{figure}

\cite{Schrijver2008} report sampling the SP data into 0.63 \arcsec pixels.
We resampled NFI observations of $B_{\rm LOS}$ at the start, middle,
and end of the 45-minute SP scanning interval from the original 0.16 \arcsec
pixels into $(4 \times 4)$ bins to achieve commensurate
resolution. Comparison of features in the $B_z$ map from SP (Figure
\ref{fig:calib_sp}, upper left) with binned $B_{\rm LOS}$ maps from
NFI at the start and end of the scan (Figure \ref{fig:calib_sp},
middle and bottom of left column)  shows that most of the discrepancies
between the SP and NFI data are in the horizontal direction: features
in the SP image extend over $N_{x,SP} = 295$ pixels in the horizontal
direction, while the NFI images of the corresponding features are only
$N_{x,NFI} = 240$ pixels in horizontal extent.

Given the position of the AR near 20$^\circ$ W from disk center at the
time of the scan, and the $\sim 10^\circ$ extent of the AR, the
cumulative effect of horizontal distortion in pixel lengths from
foreshortening in NFI pixels compared to reprojected SP pixels over
the FOV should account for a $\sim 9$\% difference, or about 22
pixels.  Hence, additional distortion is present.  We also note that,
even if reprojection fully accounted for the SP/NFI discrepancies,
un-reprojecting the SP data would require a difficult inversion of the
convolved effects of solar rotation, spacecraft pointing, and
spectrograph rastering over the 45 minute scanning period.

We therefore adopt an {\em ad hoc} approach, using repeated Fourier
interpolations to spatially interpolate the SP data onto each column
of the NFI grid.  We first interpolated the SP data onto $N_{x,NFI}$
evenly-spaced shifts $N_{x,SP}/N_{x,NFI}$ apart (one Fourier
interpolation per shift), but found systematic residual horizontal
distortions were present, consistent with a higher-order variation in
distortion across the image.  We therefore included a quadratic term
in the spacing between interpolations: where $x_i$ is the location of
the $i$-th column of NFI data, we interpolated the SP data at $x_i' =
N_{x,SP}/N_{x,NFI} (a x_i + b x_i^2)$.  By trial and error, we found
$a = 0.925$ and $b=0.06$ gave good qualitative agreement between
$\pm$100 G and $\pm$400 G contours of $B_{\rm LOS}$ from NFI at the
middle of the scan interval with $B_z$ from the SP magnetogram, as
shown in the upper right panel of Figure \ref{fig:calib_sp}.  It may
be seen that the NFI's contours of $B_{\rm LOS}$ do not match SP's
$B_z$ everywhere, but the remaining offsets do not appear to be
systematic across the image, or along columns. (We note that the
non-monotonicity in the NFI signal precludes use of some quantitative
measures of NFI-SP agreement [e.g., cross-correlation, or minimized
  square difference] since optimal alignment probably corresponds to
the {\em weakest} umbral field in NFI being co-aligned with the {\em
  strongest} umbral field in SP, while features outside of the umbra
should generally correspond one-to-one.)

As discussed above, we are primarily interested in average magnetic
properties (vertical field strength, total field strength) in
larger-scale regions of the NFI FOV, so we only use the interpolated
SP data binned into $(8 \times 8)$ and $(16 \times 16)$ macropixels.
Comparisons of $(32 \times 32)$-binned NFI $|B_{\rm LOS}|$ data with $(8
\times 8)$-binned $|B_z|$ (and $|\bvec|$) from SP in the middle-right
(and bottom-right) panel of Figure \ref{fig:calib_sp} shows good
spatial agreement at that resolution.

Note that the SP data do not fully cover the NFI FOV, which is 256$^2$
pixels when binned $(4 \times 4)$.  Consequently, we can only use the
SP data on a subset of the full NFI FOV.

%



\end{document}